\documentclass[10pt,twocolumn,twoside]{IEEEtran}

\usepackage{hyperref}
\usepackage{emptypage}  
\usepackage{cite}
\usepackage{lipsum} 
\usepackage{graphics}
\usepackage{amsmath,amssymb,bm,bbm}
\usepackage{color, subfig}
\usepackage{dsfont}
\usepackage{tikz}
\usepackage{flushend}
\usepackage{subfig,pgfplots}
\usetikzlibrary {positioning}
\def\qed{\hfill \vrule height 7pt width 7pt depth 0pt
              \medskip}
\newtheorem{thm}{Theorem}
\newtheorem{dfn}{Definition}
\newtheorem{assumption}{Assumption}
\newtheorem{prop}{Proposition}
\newtheorem{lemma}{Lemma}
\newtheorem{corollary}{Corollary}
\newtheorem{remark}{Remark}
\newcommand{\abs}[1]{\lvert#1\rvert}
\newcommand{\ds}{\displaystyle}
\newcommand{\ba}{\begin{array}}
\newcommand{\ea}{\end{array}}
\newcommand{\mb}{\boldsymbol}
\newcommand{\be}{\begin{equation}}
\newcommand{\ee}{\end{equation}}

\newcommand{\mc}{\mathcal}

\def\1{\boldsymbol{1}}
\def\0{\boldsymbol{0}}

\def\E{\mathbb{E}}
\def\R{\mathbb{R}}

\def\P{\mathbb{P}}

\def\x{{\mb x}}
\def\y{{\mb y}}
\def\u{{\mb u}}

\def \h {2.2}
\def \l {3.3}
\usetikzlibrary{patterns}
\def \ll {2.7}
\def \hh {2}
\newcommand{\rev}[1]{\textcolor{black}{#1}}

\begin{document}
%
\title{Fast Spread in Controlled Evolutionary Dynamics}
%
%
%

\author{Lorenzo~Zino,
        Giacomo~Como,~\IEEEmembership{Member,~IEEE,}
        and~Fabio~Fagnani,~\IEEEmembership{Member,~IEEE}
\thanks{Some of the results in the paper appeared in preliminary form in~\cite{Zino2017}.}
\thanks{The authors are with the  Department of Mathematical Sciences ``G.L.~Lagrange,'' Politecnico di Torino, 10129 Torino, Italy  (e-mail: {\{giacomo.como;\,fabio.fagnani\}@polito.it}).}
\thanks{Lorenzo Zino is now with the Faculty of Science and Engineering, University of Groningen,  9747 AG Groningen, The Netherlands (e-mail: lorenzo.zino@rug.nl)}
\thanks{Giacomo Como is also with the Department of Automatic Control, Lund University, BOX 118, 22100 Lund, Sweden, where he is a member of the excellence centre ELLIIT.}
\thanks{This work was partially supported by MIUR grant Dipartimenti di Eccellenza 2018--2022 [CUP: E11G18000350001], the Swedish Research Council through Project Research Grant 2015-04066, and by the Compagnia di San Paolo through a Starting Grant.}%
}


%
%

 \ifCLASSOPTIONpeerreview 
\markboth{Journal of \LaTeX\ Class Files,~Vol.~14, No.~8, August~2015}%
{Shell \MakeLowercase{\textit{et al.}}: Bare Demo of IEEEtran.cls for IEEE Journals}
%
\fi



\maketitle

\begin{abstract}
We study the spread of a novel  state in a  network, in the presence of an exogenous control. The considered controlled evolutionary dynamics is a non-homogeneous Markov process that describes the evolution of the states of all nodes in the network. Through a rigorous analysis, we \rev{estimate the performance of the system by establishing} upper and lower bounds on the expected time needed for the novel state to replace the original one. Such bounds are \rev{expressed} in terms of the support and intensity of the control policy (specifically, the set of nodes that can be controlled and its energy) and of the network topology \rev{and establish fundamental limitations on the system's performance}. Leveraging these results, we are able to classify network structures \rev{depending on the possibility to control the system using simple open-loop control policies}. Finally, we propose a feedback control policy that, using little knowledge of the network topology and of the system's evolution at a macroscopic level, allows for a substantial speed up of the spreading process \rev{with respect to simple open-loop control policies}. All these theoretical results are presented together with explanatory examples, for which Monte Carlo simulations corroborate our analytical findings.
\end{abstract}


%
\IEEEpeerreviewmaketitle

\section{Introduction}
\IEEEPARstart{I}{n} the last decades, the study of spreading dynamics in network systems has significantly advanced. On the one hand, increasingly refined models have been {proposed, with applications spanning from the spread of epidemic diseases and mutations to the diffusion of innovation and the adoption of social norms}. On the other hand, the insight into the behavior of such systems has considerably deepened, {allowing for the development of} accurate strategies to effectively {control and influence}  the spreading process. The literature on epidemics offers a paradigmatic example, where the understanding of how the network of interactions between the individuals influences the spread of a disease~{
\cite{Ganesh2005,Draief2006,Pastor-Satorras2015,Mei2017,Fagnani2019}} has paved the way for the study of control policies to mitigate or stop epidemic outbreaks~{\cite{Chung2009,Borgs2010,Drakopoulos2014, Trajanovski2015,Nowzari2016}.}
Also, for {decision-making in social systems}, the extensive analysis of the voter model~\cite{Ligget1985,Frachebourg1996} and { the study of dynamical interaction models in game theoretic frameworks~\cite{Montanari2010,Young2011,Rossi.ea:2019}} have allowed for the design of techniques to optimally place a spreader in a network~\cite{Kempe2003,Durand.ea:2020} and to control networks of imitative agents~\cite{Riehl2017}.

In this paper, we focus on a dynamical network system that models the spread of {novel states} inside a population. It first appeared in the literature in~\cite{Lieberman2005} under the name of \emph{evolutionary dynamics}. {These dynamical systems have been used to model the evolutionary competition between a novel species and the original one in a geographic region.} 
In evolutionary dynamics, the geographical pattern is modeled as a {weighted} network whose nodes {represent} regions, links represent proximity, {and the links' weights measure the level of interaction between individuals in adjacent regions.} Each node can be in two possible states, representing the original and the novel species, respectively. The spreading process follows a probabilistic rule and takes place through pairwise interactions between {adjacent nodes, which yield a competition between the species present in the two nodes to place their offsprings}. 
This model naturally accounts for possible stochastic biases that favor one specie against the other in the pairwise competitions, thus reflecting an evolutionary advantage of one specie with respect to the other one. The key problems considered in the literature {of evolutionary dynamics~\cite{Lieberman2005,Rychtar2008,Ohtsuki2006,Broom2011,Allen2017} focus on the understanding of} how the network structure influences the probability that the mutant state (typically with an evolutionary advantage) diffuses in the network ---that is often referred to as the fixation probability--- and the duration of such a spreading process. 

This model {may also find valuable applications in other fields, e.g., to study the diffusion of innovation in social systems. In this setting, the two states may represent two alternative technologies, and the nodes represent users interacting in a pairwise fashion on a social network and exchanging information on the used products. As a result of these pairwise interactions, one of the two individuals may get convinced to adopt the technology used by the other one, thus leading to an imitative/contagion mechanism. As before, the model can easily incorporate an intrinsic bias towards one technology with respect to the other, thus accounting, e.g., for an intrinsic quality difference between the two products.}

Despite its importance, few analytical results for evolutionary dynamics are available in the literature. To the best of our knowledge, the fixation probability has been analytically computed for very specific network topologies~\cite{Rychtar2008,Ohtsuki2006}, while most of the results are based on extensive Monte Carlo simulations~\cite{Lieberman2005,Broom2011,Allen2017} and no effective policy to control the spreading in the framework of evolutionary dynamics has been studied.
In a preliminary work~\cite{Zino2017}, we have proposed a new formalism for evolutionary dynamics, which presents two novelties with respect to the original one: i) the spreading process is modeled through a link-based (instead of a node-based as in~\cite{Lieberman2005}) activation mechanism; and ii) an exogenous control action is incorporated {to model the controlled introduction of the novel species in a geographic region}. This change of perspective and the explicit introduction of a control action have enabled us to gain new analytical insight concerning the expected duration of the spreading process.
In~\cite{Zino2017}, we introduced a first feedback control policy to speed up the spreading process, but its feasibility was limited to very specific networks {and it} presented two drawbacks: it was extremely sensible to small data errors and very costly from the control viewpoint. {An improved} feedback control policy was proposed in \cite{iscas2018} and there tested on a real-world case study by means of Monte Carlo numerical simulations.

In this paper, we undertake a fundamental analysis of controlled evolutionary network system in great generality, encompassing both open-loop and feedback control policies. Control actions are evaluated on the basis of their cost ---measured in terms of both the number of nodes where the control acts and of its total energy--- and the spreading time of the induced controlled evolutionary network systems. {Our main contribution is the establishment of fundamental limitations between these quantities. Specifically, Theorem \ref{teo:main upper} and Corollary \ref{cor:fund} state general bounds on the expected spreading time in terms of the control policy implemented, its cost, and the topological network properties: notably, we highlight the key role played by the network expansiveness. From these general results, we derive an array of ready-to-use tools to estimate these quantities for specific control policies and network structures.} In particular, this allows for performing a full theoretical analysis of the feedback control law presented in~\cite{iscas2018}, assessing its performance.


{From a technical viewpoint, our study is based on the estimation of the expected time needed for the Markov process modeling the controlled network evolutionary dynamics to reach its absorbing state. The presence of the external control makes this Markov process non-homogeneous and hinders a direct application of established results on the convergence time of homogeneous Markov processes~\cite{Mihail1989,Roberts1996}. Our analysis is based on two key observations. The first one is that the considered controlled evolutionary network systems present a natural monotonicity property with respect to the initial configuration of nodes in the novel state as well with respect to the parameter measuring the evolutionary advantage of the novel state. Specifically, such monotonicity property can be formalized in terms of stochastic domination for the induced Markov process. The second observation is that, as in most epidemic models based on contact processes~\cite{Drakopoulos2014}, the speed of diffusion of the novel state is proportional to the size of the boundary  separating the set of nodes in the novel state from that of nodes in the original one in the network. When the network is an expansive graph, whereby the size of the boundary of subsets of nodes is proportional to their cardinality, this creates a positive feedback for the considered network evolutionary dynamics that is shown to imply fast spread (in a time that is logarithmic in the network size), even in the absence of an external control (except for the initial seeding). On the other hand, in non-expansive networks, the graph topology typically prevents this positive feedback, as the system may reach configurations corresponding to bottlenecks where the spreading process is significantly slowed down. It is in these configurations that an external control is the most valuable in reinforcing the spreading process. In fact, the feedback control policies analyzed in Section \ref{sec:feedback} are based exactly on the idea of concentrating the control efforts when the system has entered configurations whose boundary is too small to guarantee a fast spread. }


{To better illustrate the implications of our general results, we present the rigorous analysis of three fundamental network examples: expander graphs (encompassing many models of social networks such as small-world~\cite{Watts1998} and scale-free networks~\cite{Barabasi2002}), stochastic block models (SBMs)~\cite{Holland1983} and ring graphs. Expander and ring graphs are presented as extreme instances of highly and poorly connected structures, respectively, and constitute benchmarks for fast and slow spread behaviors, respectively. On the other hand, SBM is an established model to represent geographic regions and real-world clustered communities~\cite{Fortunato2016} and it exhibits a quite interesting behavior. Our theoretical results show that, for SBMs, fast diffusion can not be reached using open-loop control policies,  it can instead be achieved using the proposed feedback control law. Finally, Monte Carlo numerical simulations are offered to corroborate our theoretical results on these three examples.} We believe that the capability {of incorporating and analyzing} control architectures in spreading dynamics and, more generally, in all types of network dynamics is a crucial step from the application viewpoint. 

{The rest of this paper is organized as follows. In Section~\ref{section model}, we introduce the  model. In Section~\ref{sec:analysis}, we present the main technical results. Section~\ref{sec:constant} is devoted to a detailed analysis of open-loop control policy, in particular constant control policies, for the three fundamental network examples. In Section~\ref{sec:feedback}, we propose and analyze a feedback control policy, proving its effectiveness for SBMs. Section~\ref{section conclusions} summarizes the work and outlines future research. The paper ends with an Appendix containing the proofs of the most technically involved results.}

\subsection{Notation}

We gather here some notational conventions followed throughout the paper. {The $n$-dimensional all-$0$ and  all-$1$ vectors are denoted by $\0$ and $\1$, respectively},  $\delta^{(i)}$ denotes a vector with a $1$ in the $i$th entry and $0$ otherwise, and, for every subset {$\mc S\subseteq\{1,\dots,n\}$}, $\delta^{(\mc S)}=\sum_{i\in \mc S}\delta^{(i)}$. {We denote by $\R_{+}$} the set of non-negative real numbers. 
Given a vector $\x\in\R^n$, $\x^\top$ denotes its transpose. Given two vectors $\x,\y\in\R^n$, $\x\geq \y$ denotes entrywise inequality, i.e., $x_i\geq y_i,$ for {all} $i\in\{1,\dots, n\}$. {By ${\mathbbm 1}_{A}$ we denote} the indicator function of set $A$. {By convention, the value of an empty sum is equal to $0$ and the ratio between a positive number and $0$ is $+\infty$.  Finally, the left- and right-hand limits of a function $f$ in $t_0$ are denoted by $$f(t_0^-):=\lim_{t\nearrow t_0}f(t)\,,\quad f(t_0^+):=\lim_{t\searrow t_0}f(t)\,.$$ 
respectively.}

\section{The Model}\label{section model}

In our model, the network is described as a weighted undirected graph whose nodes represent agents that are either in state $0$ (original state) or in state $1$ (novel state). A spreading mechanism and an exogenous control determine the evolution of  the states of the system over time.\smallskip

{\bf Weighted undirected graph.} We consider a connected graph $\mc{G}=( \mc{V}, \mc{E},W)$ with node set $\mc V=\{1,\dots,n\}$, undirected link set $\mc E$, {and symmetric weight matrix $W\in\R_{+}^{n\times n}$, whose positive entries $W_{ij}=W_{ji}>0$ correspond to the weights of the links $\{i,j\}\in\mc E$.}\smallskip

{\bf Spreading mechanism.} Each undirected link $\{i,j\}$ of the graph $\mc G$ is equipped with an independent rate-$W_{ij}$ Poisson clock modeling the interactions between nodes $i$ and $j$. When the clock associated with  link $\{i,j\}$ ticks, if both nodes $i$ and $j$ are in the same state, nothing happens. Otherwise, if the current states of nodes $i$ and $j$ differ from one another, a conflict takes place and the winning state occupies both nodes. Conflicts are solved in a probabilistic way: the novel state $1$ wins a conflict (independently of the others) with a fixed probability $\beta$, while the original state $0$ wins with probability $1-\beta$. {The parameter $\beta$ in $[0,1]$ captures the \emph{evolutionary advantage} of the original state (if $\beta<1/2$) or of the novel state (if $\beta>1/2$) in the considered network dynamics.} \smallskip

{\bf Exogenous control.} We fix a locally integrable function $U:\R_{+}\to\R^n_{+}$ whose $i$-th entry $U_i(t)$ represents the control rate at which the novel state is enforced at node $i$ at time $t$. \smallskip

The triple $(\mc G, \beta, U(t))$ shall be referred to as the \emph{controlled evolutionary dynamics}. To it, we now associate a Markov process $X(t)$ that describes the spread of the novel state in the network~\cite{Levin2009}. {The process $X(t)$ takes values in the configuration space $\{0,1\}^n$ and has the following interpretation: $$X_i(t)=\left\{\begin{array}{ll}0&\text{if at time $t$ node $i$ is in the original state,}\\
1&\text{if at time $t$ node $i$ is in the novel state.}\end{array}\right.$$ 
The evolution of $X(t)$ is described as a time non-homogeneous jump Markov process where the only transitions that can possibly take place from $X(t)=\x$ are towards configurations that differ from $\x$ in a single entry. Specifically, 
given a configuration $\x$ and a node $i$, we consider the transition rates from $\x$ to $\x+ \delta^{(i)}$ (if $x_i=0$) and to $\x- \delta^{(i)}$ (if $x_i=1$) at time $t$ formally defined as
$$
\lambda_i^\pm(\x,t):=\lim_{h\searrow0}\frac1h\P[X(t+h)=\x\pm \delta^{(i)}|X(t)=\x],
$$
as illustrated in Figure~\ref{fig:markov} (see~\cite{Levin2009} for more details on Markov processes). Our model is then completely described by putting}
\begin{equation}\label{eq:transitions}\left.\begin{array}{llll}
\lambda_i^+(\x,t)&=&(1-x_i)\big(\beta(W \x)_i+U_i(t)\big)\\[4pt]
\lambda_i^-(\x,t)&=&x_i(1-\beta)\left(W(\1-\x)\right)_i\,.
\end{array} \right.
\end{equation}
{{Notice how the network topology and the exogenous control determine which transitions can actually take place and their corresponding rates. Indeed, the rate of transition with which the state of node $i$ can jump from $0$ to $1$ at time $t$ is the sum of two terms: a first term $\beta(W \x)_i$ that is proportional to the total weight of links pointing towards neighbors of node $i$ that are already in state $1$ at time $t$; and a second term $U_i(t)$ given by the control effort exerted on node $i$ at time $t$.
On the other hand, for the same node $i$, the rate of transition from $1$ to $0$ is determined by the term $(1-\beta)\left(W(\1-\x)\right)_i$ that is proportional to the total weight of links pointing towards neighbors of node $i$ that are already in state $0$ at time $t$.}  


\begin{figure}
\begin {tikzpicture}
  \node[draw, circle, fill=white] (1) at (0,-.5) {};
    \node[draw, circle] (2) at (1,-.5) {};
        \node[draw, circle] (3) at (-1,-.5) {};
            \node[draw, circle,fill=red] (4) at (0,-1.5) {};
                \node (5) at (0,.5) {};
                \node[draw, circle,fill=red] (6) at (1,-1.5) {};
\node(7) at (-1.6,-.5) {};
   \node at (0,-.5) {\footnotesize{$i$}};
\node(8) at (1.5,-.4) {};
\node(9) at (3,-.4) {};
\node(10) at (1.5,-.6) {};
\node(11) at (3,-.6) {};
\draw[->,thick, >=latex] (8) edge [bend left] node[above] {{$\lambda_i^+(\x)$}} (9);
\draw[->,thick, >=latex] (11) edge [bend left] node[below] {{$\lambda_i^-(\x)$}} (10);

\draw[-, >=latex] (1) edge (2);
\draw[-, >=latex] (1) edge (3);
\draw[-, >=latex] (1) edge (4);
\draw[dashed, >=latex] (1) edge (5);
\draw[-, >=latex] (6) edge (4);
\draw[dashed, >=latex] (6) edge (1.6,-1.5);
\draw[dashed, >=latex] (3) edge (-1,-1.5);
\draw[dashed, >=latex] (7) edge (3);
\end{tikzpicture}\,\,\begin {tikzpicture}
  \node[draw, circle, fill=red] (1) at (0,-.5) {};
    \node[draw, circle] (2) at (1,-.5) {};
        \node[draw, circle] (3) at (-1,-.5) {};
            \node[draw, circle,fill=red] (4) at (0,-1.5) {};
                \node (5) at (0,.5) {};
                \node[draw, circle,fill=red] (6) at (1,-1.5) {};
\node(7) at (-1.6,-.5) {};
   \node at (0,-.5) {\footnotesize{${i}$}};

\draw[-, >=latex] (1) edge (2);
\draw[-, >=latex] (1) edge (3);
\draw[-, >=latex] (1) edge (4);
\draw[-, >=latex] (6) edge (4);
\draw[dashed, >=latex] (6) edge (1.6,-1.5);
\draw[dashed, >=latex] (3) edge (-1,-1.5);

\draw[dashed, >=latex] (1) edge (5);
\draw[dashed, >=latex] (7) edge (3);
\end{tikzpicture}
\caption{Transitions of the jump Markov process $X(t)$.}
\label{fig:markov}\end{figure}
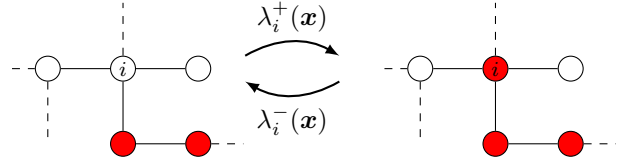

\begin{remark}
In the absence of the external control (i.e., if $U(t)=\0$ for every $t\geq 0$), the model reduces to a voter model~\cite{Ligget1985}, which is biased if $\beta\neq 1/2$~\cite{Ferreira1990}. A homogeneous version of our model without control and where clocks are associated with nodes instead of links, has been proposed  in~\cite{Lieberman2005}. However, the analytical results on such a model are limited to the computation or the estimation of the probability that the novel state diffuses to the whole network (named fixation probability), while more in-depth analysis is limited to specific network structures~\cite{Rychtar2008}--\cite{Ohtsuki2006}, and the convergence time is studied only through Monte Carlo simulations.
\end{remark}

\begin{remark}
{An equivalent model can be obtained if we replace the exogenous controller with a fictitious stubborn node $s$ having fixed state $X_s(t)=1$, for $t\geq 0$ and we add time-varying links from every other node $i$ to $s$ having weights $W_{is}=U_i(t)/\beta$. In particular, if $U_i(t)=0$ we assume that there is no link between $i$ and $s$ at time $t$. The dynamics driven by the spreading mechanism on this (time-varying) graph leads to exactly the same Markov process. In opinion dynamics, models with stubborn nodes have been analyzed~\cite{Acemoglu2013,Como.Fagnani:2016,proskurnikov2017tutorial}. These models typically consider the presence of multiple stubborn nodes having possibly different states but assume the graph to be constant and the main problem addressed concerns the way the various stubborn nodes influence the asymptotic behavior of the system. The problem considered in this paper concerns instead the transient behavior of the system and its analysis requires different techniques.}
\end{remark}

Throughout the paper, we make the following assumptions.

\begin{assumption}\label{ass:beta} The probability that in a conflict the novel state wins over the original one is $\beta>1/2$.
\end{assumption}

\begin{assumption}\label{ass:control} If $X(t)=\0$, then $U(t)\neq \0$.
\end{assumption}

The first one captures an evolutionary advantage of the novel state and the second one the presence of an exogenous control to force the evolutionary dynamics whenever the system is in the novel-free configuration $\0$.

{From the structure of the transition rates~\eqref{eq:transitions}, Assumption~\ref{ass:control}, and the fact that the graph $\mc G$ is undirected and connected}, it follows that the all-$1$ configuration $\1$ is the only absorbing state of the system and it is reachable from every other state. Hence, the novel state eventually spreads to the whole network almost surely in finite time~\cite{Levin2009}. From an application perspective, our interest is to {shed light} on the transient behavior of the system. To this aim, we introduce the following two performance indices:
\begin{itemize}
\item the \emph{spreading time}, that is,
$$
T:=\inf\left\{t\ge 0:X(t)=\1\right\}\,;
$$
\item the \emph{control cost}, that is,
$$
J=\int_0^T \1^\top U(t)\text{dt}\,.
$$
\end{itemize} 
{In this paper, we will mostly focus on the estimation of the expected values of these two indices. Since these may depend on the initial configuration of the system $X(0)$, we shall use the following notation for the conditional expected values  of the spreading time and of the control cost, respectively:  
$$\begin{array}{l}\E_{\x_0}[T]: =\E[T\,|\,X(0)=\x_0]\,,\\ \E_{\x_0}[J]: =\E[J\,|\,X(0)=\x_0]\,.\end{array}$$
For the purpose of this paper, we will typically be interested in the scenario with $X(0)=\0$, that is the novel-free configuration. Hence, most of our results will be expressed as bounds on $\E_{\0}[T]$ and $\E_{\0}[J]$.
}

We will consider two main types of control policy, described in the following.\smallskip

{\bf Open-loop control policies}, where the control signal $U(t)$ is predetermined. A simple example of such control policies are those with $U(t)=\u$ constant for every $t\geq 0$. We will refer to these as \emph{constant control policies}. \smallskip

{\bf Feedback control policies}, where the control signal $U(t)$ is chosen as a function of the process $X(t)$ itself. Precisely, in this case we consider a function {$\nu:\{0,1\}^n\to \R^n_{+}$} and we take $U(t)=\nu(X(t))$. The triple $(\mc G, \beta, \nu)$  is then called a \emph{feedback controlled evolutionary dynamics}. In this case, defining the transition rates of the process $X(t)$ as in \eqref{eq:transitions}, we have that $X(t)$ is a time-homogeneous jump Markov process.\smallskip

There are typically some constraints that we want to enforce on the admissible control policies. In general, the external control $U(t)$ is constrained to be active only at certain specified nodes, denoted by $\mc U\subseteq \mc V$. In this case, the triple $(\mc G, \beta, U(t))$ is called an \emph{$\mc U$-controlled evolutionary dynamics}.

We introduce here some fundamental quantities that will be used in the following. 

\begin{dfn} Consider an undirected weighted graph $\mc G=(\mc V,\mc E,W)$, with $\abs{ \mc V}=n$. 
\begin{itemize}
\item 

Given a subset of nodes $\mc W\subseteq\mc V$, we define its \emph{weighted boundary}~\cite{Goel2006} as
\begin{equation}\label{weighted boundary}
{\zeta(\mc W)}=\sum_{i\in \mc W}\sum_{j\notin \mc W}W_{ij}.
\end{equation}
Since $W$ is symmetric, then {$\zeta(\mc  W)=\zeta(\mc V\smallsetminus \mc W)$}. {Since $\mc G$ is connected, then $\zeta(\mc W)>0$, for all $\mc W\notin\{\mc V,\emptyset\}$.} 

\item The \emph{minimum conductance profile} of $\mc G$ is a function {$\phi:\{1,\dots,n-1\}\to\R_{+}$}, defined as
\begin{equation}\label{eq:min cond}
\phi(a)=\min_{\mc W\subset  \mc V,\abs{\mc W}=a}{{\zeta(\mc W)}}.
\end{equation}

\item The \emph{maximum expansiveness profile} is a function {$\eta:\{1,\dots,n-1\}\to\R_{+}$}, defined as
\begin{equation}\label{eq:max exp}
\eta(a)=\max_{\mc W\subset  \mc V,\abs{\mc W}=a}{{\zeta(\mc W)}}.
\end{equation}
\end{itemize}
\end{dfn}

 {We also define three one-dimensional stochastic processes as aggregate statistics of the process $X(t)$:}
\begin{itemize}
\item  the \emph{number of nodes} with the novel state
$$A(t):=\1^\top X(t)\,;$$
 \item the size of the \emph{boundary} of the set of nodes that are in the novel state
$$
B(t):=
X(t)^\top W(\1-X(t))\,;\,\text{and}$$
 \item  the \emph{effective control rate} in nodes that are in the original state
$$C(t):=\left(\1-X(t)\right)^\top U(t)\,.$$
\end{itemize}


\section{General Performance Analysis}\label{sec:analysis}

In this section, we present a series of theoretical results that allow one to estimate the performance of controlled evolutionary dynamics. First, we focus on results guaranteeing the expected spreading time to be below a certain bound that depends on the network topology and the form and intensity of the control effort. Then, we present an array of results illustrating fundamental limitations of the controlled evolutionary { dynamics}. They are expressed in the form of control-independent lower bounds on the expected spreading time.  {Leveraging these general results, ready-to-use corollaries will be derived for specific choices of the control policy.}

\subsection{Performance Guarantees}\label{sec:upper}
{In this subsection, we present a general result estimating the performance of controlled evolutionary dynamics, which will be then applied to both open-loop and feedback control policies.}

{The core of the proposed analysis consists in the estimation of the expected time spent by the process $X(t)$  in the various {non-absorbing} configurations before reaching the all-$1$ absorbing configuration. To perform such an analysis, we focus on the aggregate process $A(t)=\1^\top X(t)$ counting the number of state-$1$ nodes in the current configuration, hence taking values in $\{0,1,\ldots,n\}$, and estimate the time spent by $A(t)$ before being absorbed in state $n$. The analysis of this process is nontrivial because, in general, $A(t)$ is not a Markov process. Indeed, notice that each transition of $X(t)$ increases or decreases $A(t)$ by $1$ unit and that, by \eqref{eq:transitions}, in general such transitions depend on the complete vector $X(t)$ through the network weight matrix $W$ and not just on the aggregate variable $A(t)$. From \eqref{eq:transitions}, we can compute the transition rates of the process $A(t)$ conditioned on $X(t)$. Precisely, when $A(t)=a$, its rate of increase (corresponding to the transition to $a+1$) conditioned on $X(t)=\x$ is equal to}
\be\label{trans+}
\begin{array}{lrl}
\lambda_t^+(a|\x)&:=&\ds\lim_{h\searrow 0}\frac{1}{h}\mathbb P[A(t+h)=a+1|X(t)=\x]\\[10pt]
&=&\ds\sum_{i\in \mc V}\lambda_i^+(\x,t)\\[10pt]
&=&\ds\beta(\1-\x)^\top W \x+\left(\1-X(t)\right)^\top U(t)\\[10pt]
&=&\ds\beta B(t)+C(t).
\end{array}
\ee
Notice that this quantity depends on {the configuration $X(t)$ only through the two scalar aggregate statistics} $B(t)$ and $C(t)$. {Similarly, the rate of decrease of $A(t)$ is equal to}
\be\label{trans-}
\ba{rcl}\lambda_t^-(a|\x)&=&\ds\frac1h\lim_{h\searrow 0}\mathbb P[A(t+h)=a-1|X(t)=\x]\\[10pt]
&=&(1-\beta) B(t),\ea
\ee
and this quantity depends {on the configuration $X(t)$ only through the size of the boundary} $B(t)$. The transitions of the aggregate process $A(t)$ {and their corresponding rates are shown} in Figure \ref{z process}.

\begin{figure}
\center
\begin {tikzpicture}
\def \s {1.5}
\node[draw, circle] (0) at (0,0) {$0$};
\node[draw, circle] (1) at (\s,0) {$1$};
\node[draw=none, circle] (2) at (2*\s,0) {$\dots$};
\node[draw, circle,text width=0.5em,text centered] (3) at (3*\s,0) {$a$};
\node[draw=none, circle] (4) at (4*\s,0) {$\dots$};
\node[draw, circle] (5) at (5*\s,0) {$n$};

\path [->, >=latex]  (0) edge[bend left =20]   node [above] {\small{$C(t)$}} (1);
\path [->, >=latex]  (1) edge[bend left =20]   node [above] {\small{$\lambda^+(a|\x)$}} (2);
\path [->, >=latex]  (2) edge[bend left =20]   node [above] {\small{$\lambda^+(a|\x)$}} (3);
\path [->, >=latex]  (3) edge[bend left =20]   node [above] {\small{$\lambda^+(a|\x)$}} (4);
\path [->, >=latex]  (4) edge[bend left =20]   node [above] {\small{$\lambda^+(a|\x)$}} (5);
\path [->, >=latex]  (2) edge[bend left =20]   node [below] {\small{$\lambda^-(a|\x)$}} (1);
\path [->, >=latex]  (3) edge[bend left =20]   node [below] {\small{$\lambda^-(a|\x)$}} (2);
\path [->, >=latex]  (4) edge[bend left =20]   node [below] {\small{$\lambda^-(a|\x)$}} (3);
\path [->, >=latex]  (1) edge[bend left =20]   node [below] {\small{$\lambda^-(a|\x)$}} (0);
\end{tikzpicture} 
\caption{Transition graph of the continuous-time process $A(t)$.}
\label{z process}
\end{figure}
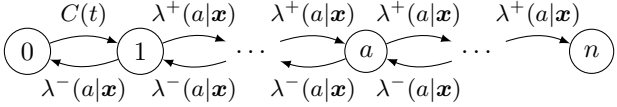

{In the following, it will be useful to consider the transition probabilities associated to these transition rates~\cite{Levin2009}. Precisely, for all $t\ge0$, $a\in\{1,\dots,n-1\}$ and $\x\in\{0,1\}^n$ let $p_t^+(a|\x)$ ($p_t^-(a|\x)$) be the conditional probability that, should the process $X(t)$ have a state transition at time $t$,  such transition would increase the value of $A(t)$ from $a$ to $a+1$ (respectively, decrease the value of $A(t)$ from $a$ to $a-1$), given the current configuration $X(t^-)=\x$.} We have
\be\label{eq:l1}
p_t^+(a|\x)=\ds\frac{\lambda_t^+(a|\x)}{\lambda_t^+(a|\x)+\lambda_t^-(a|\x)}
=\frac{\beta B(t)+ C(t)}{B(t)+C(t)}\geq \beta\,,\ee
\be\label{eq:l2}
p_t^-(a|\x)=\ds\frac{\lambda_t^-(a|\x)}{\lambda_t^+(a|\x)+\lambda_t^-(a|\x)}=\frac{(1-\beta)B(t)}{B(t)+C(t)}\leq 1-\beta\,,
\ee
{i.e., these two probabilities can be uniformly bounded by functions that are independent of the state $\mb x$ and time $t$, and only depend on the evolutionary advantage $\beta$. For $a=0$, such transition probabilities reduce to $p_t^+(0|{\0})=1$ and $p_t^-(0|{\0})=0$, for every $t\geq 0$.}

{In our theoretical analysis, we need to bound the conditional probability that, given that the system is in a given configuration $X(t_0)=\x$ with $\1^\top\x=a$ state-$1$ nodes at time $t_0\geq 0$, the aggregate process $A(t)$ will ever go below $a$ at some time $t>t_0$, before being absorbed into its absorbing state $n$. Since the transition rates of the aggregate process $A(t)$ depend on the whole non-homogeneous Markov process $X(t)$, such probability may in general depend of the configuration of the system $X(t_0)$ and on the time $t_0$. In our derivations, we will need a lower bound on such a quantity that is uniform with respect to these two variables. To this aim, we define the following lower bound on such a probability:
$$
q_a:=\inf_{t_0\geq 0}\,\,{\min\limits_{\x\, :\, \1^\top \x=a}} \P[A(t)\geq a,\, \forall t\geq t_0\, |\, X(t_0)=\x ]\,.
$$
Such a uniform bound is estimated using \eqref{eq:l1} in the following result. Its proof uses standard coupling arguments between stochastic processes and, for the sake of making the paper self-contained, is reported in Appendix~\ref{section analysis}. 

\begin{lemma}\label{prop:Moran}
For every $a\in\{1,\dots,n-1\}$, $
q_a\geq{(2\beta-1)}/{\beta}\,.$
\end{lemma}}

{Observe that Lemma \ref{prop:Moran} provides a  lower bound on $q_a$ and, ultimately, on the desired probability,  depending only on the evolutionary advantage $\beta$. Using this uniform bound, we can now estimate the total amount of time that the process spends in each of its non-absorbing states, when a relation between the three aggregate statistics $A(t)$, $B(t)$, and $C(t)$ is verified. Specifically, the following is the key technical result which allows us to finally formulate our performance guarantees.}

\rev{\begin{prop}\label{prop:time} Let $T_a$ be the random variable measuring the total amount of time spent by process $A(t)$ in state $a$ before being absorbed in $n$. Then, for every  $f:\{0,\dots,n-1\}\to\R_{+}$ such that
\be\label{eq:assumption upper_lemma} B(t)+C(t)\geq f(A(t)), \ee 
 for every $t\geq 0$ such that $X(t)\neq \1$, the conditional expected spreading time satisfies
\be\label{recursive-time-gen} \E[T_a\,|\,X(0)=\0]\leq \left\{\ba{ll} \ds\frac{\beta}{2\beta -1}\frac{1}{f(0)}\quad &{\rm if}\, a=0\,,\\[8pt]
\ds\frac{1}{2\beta -1}\frac{1}{f(a)} \quad &{\rm if}\, a\neq 0\,.\ea\right.\ee
\end{prop}}

\begin{IEEEproof}{
For $s\ge0$ and $a\in\{0,1,\ldots,n-1\}$, let $T_a(s)$ denote the time spent by the process $A(t)$ in state $a$ {from time $s$ on.} Clearly, $T_a=T_a(0)$ and, for every $s\ge0$ and $\x_0\in\{0,1\}^n$, 
$$\E[T_a(s)\,|\,X(s)=\x_0,X(0)=\0]=
\E[T_a(s)\,|\,X(s)=\x_0]\,,$$
because of the Markov property. For $a\in\{0,\dots,n-1\}$, put
\be\label{theta_a}\theta_a:=\sup_{s\geq 0}\,\max_{\x_0:\1^\top \x_0\leq a}\E[T_a(s)\,|\,X(s)=\x_0]\,.\ee
For $t\ge0$, {we define the random variable} 
$$S(t)=\inf\{h\ge 0:X\big((t+h)^+\big)\ne X(t^+)\}\,,$$
{which models the waiting time for the first configuration change after time $t$. {To estimate its expected value conditioned to the event $X(t^+)=\x$, we observe that, for $r\in ]t, t+S(t)]$, the total jump rate of the process $A(t)$ can be estimated, using \eqref{trans+} and \eqref{trans-} and the relation in \eqref{eq:assumption upper_lemma}, as
$$\xi(r)=\lambda_r^+(\1^\top\x|\x)+\lambda_r^-(\1^\top\x|\x)=B(r)+C(r)\geq f(\1^\top \x)\,$$
Standard properties of non-homogeneous Markov jump processes then yield
$$\begin{array}{rcl}\E[S(t)|X(t^+)=\x]&=&\ds\int_{0}^{+\infty}e^{-\ds\int_t^{t+s} \xi(r)\text{d}r}\text{d}s\\[7pt]
&\leq&\ds\int_{0}^{+\infty}e^{-\ds\int_t^{t+s}f(\1^\top \x)\text{d}r}\text{d}s\\[7pt]&=&\ds\int_{0}^{+\infty}e^{-sf(\1^\top \x)}\text{d}s\\[7pt]
&\leq&\ds\frac1{f(\1^\top \x)}\,.\end{array}$$}}}


We fix an initial time $t_0$ (not necessarily equal to $0$) and an initial state $X(t_0)=\x_0$ such that $\1^\top \x_0\leq a$. 
Let $t_1\geq t_0$ be the time when $A(t)$ reaches the value $a$ for the first time after $t_0$. {Then, using Markov property and separating the time spent in $a$ during its first entrance from the other contributions to $T_a(t_0)$, we write the following recursive expression:}
\be\label{recursive-time}
\ba{l}
\E[T_a(t_0) |X(t_0)= \x_0, X(t_1^+)=\x_1 ]\\[5pt]
\,=\E[T_a(t_1) | X(t_1^+)=\x_1 ]\\[5pt]
\,=\E[S(t_1)| X(t_1^+)=\x_1 ]+\E[T_a(t_1+S(t_1))|X(t^+_1)=\x_1)].\ea\ee
{Let us define $t_2:=t_1+S(t_1)$ and $\x_2:= X(t_2^+)$. Then, we estimate $\E[T_a(t_2)|X(t^+_1)=\x_1)]$ depending on the direction of the jump that has occurred at time $t_2$. If $A(t)$ has moved backward, then $\1^\top\x_2=a-1$ and, by the definition in \eqref{theta_a}, 
\be\label{recursive-time_p1}\E[T_a(t_2)\, |\, \1^\top\x_2=a-1]\leq \theta_a.\ee
If $A(t)$ instead has increased to $a+1$, the probability that the process $A(t)$ will return to $a$ before getting absorbed in $n$ is bounded from the above by $(1-q_{a+1})$, yielding 
\be\label{recursive-time_p2}\E[T_a(t_2)\,|\,\1^\top\x_2=a+1]\leq (1-q_{a+1})\theta_a.\ee 
Inserting \eqref{recursive-time_p1} and  \eqref{recursive-time_p2} into \eqref{recursive-time},} we bound
\be\label{recursive-time1}\ba{l}\E[T_a(t_0) \, |\, X(t_0)=\x_0, X(t_1^+)=\x_1 ]\\[2pt]
\quad\leq \ds \frac{1}{f(a)}+p_{t_2^-}^-(a|\x_1)\theta_a+p_{t_2^-}^+(a|\x_1)(1-q_{a+1})\theta_a.\ea\ee
For $a=0$, we have $p_{t_2^-}^-(0|\x_1)=0$ and $p_{t_2^-}^+(0|\x_1)=1$. Hence, using Lemma \ref{prop:Moran} we obtain
$$\E[T_0(t_0)|X(t_0)=\0]\le\frac{1}{f(0)}+(1-f_1)\theta_0
\leq \frac{1}{f(0)}+\frac{1-\beta}{\beta}\theta_0\,.$$
{This holds true for every $t_0\geq 0$ and thus, by taking the supremum of the left-hand side with respect to $t_0$, we obtain} 
$$\theta_0\leq \frac{1}{f(0)}+\frac{1-\beta}{\beta}\theta_0\,,$$
so that
\be\label{recursive-time2}\theta_0\leq \frac{\beta}{2\beta -1}\frac{1}{f(0)}.\ee
For $a\in\{1.\ldots,n-1\}$, applying \eqref{eq:l1} and \eqref{eq:l2} to the right-hand side of \eqref{recursive-time1}, we obtain
\be\label{recursive-time3}\ba{l}\E[T_a(t_0)\,|\, X(t_0)=\x_0, X(t_1^+)=\x_1 ]\\[5pt]
\quad\leq \ds \frac{1}{f(a)}+(1-\beta)\theta_a+\beta\frac{1-\beta}{\beta}\theta_a\\
\quad\ds=\frac{1}{f(a)}+2(1-\beta)\theta_a.\ea\ee
Since this holds true for every $t_0\ge0$ and $\x_0\in\{0,1\}^n$ such that $\1^\top \x_0\leq a$, we immediately obtain that
$$\theta_a\leq \frac{1}{f(a)}+2(1-\beta)\theta_a,$$
yielding
\be\label{recursive-time4} \theta_a\leq \frac{1}{2\beta -1}\frac{1}{f(a)}.\ee
{Since $\mathbb  E\left[T_a|X(0)=\0\right]\leq\theta_a$, the claim now follows from \eqref{recursive-time2} and \eqref{recursive-time4}.}
\end{IEEEproof}\medskip

We can now present and prove the fundamental result of this section.

\begin{thm}\label{teo:main upper}
Let $(\mc G, \beta, U(t))$ be a controlled evolutionary dynamics {with initial configuration $X(0)={\0}$}.
Then, for every $f:\{0,\dots,n-1\}\to\R_{+}$ such that 
\be\label{eq:assumption upper} B(t)+C(t)\geq f(A(t))\,,\ee  
for $t\ge0$, {the conditional expected spreading time}  satisfies
{\begin{equation}\label{bound F}
\E_{\0}[T]\leq\frac{\beta}{(2\beta-1)f(0)}+\frac{1}{2\beta-1}\sum_{a=1}^{n-1}\frac{1}{f(a)}\,. 
\end{equation}}
\end{thm}
\begin{IEEEproof}
It follows from Proposition \ref{prop:time} that
$$
\ba{lll}
\E_{\0}[T]
&=&\ds\E\left[\sum_{a=0}^{n-1}T_a\,\Big|\,X(0)=\0\right]\\[7pt]
&=&\ds\mathbb  E\left[T_0\,|\,X(0)=\0\right]+\sum_{a=1}^{n-1}\mathbb E\left[T_a\,|\,X(0)=\0\right]\\[7pt]
&\leq&\ds\frac{\beta}{(2\beta-1)f(0)}+\frac{1}{2\beta-1}\sum_{a=1}^{n-1}\frac{1}{f(a)}\,,
\ea
$$
thus proving the claim. 
\end{IEEEproof}\medskip

\rev{Notice how the quantity $B(t)+C(t)$ appearing in the left-hand side of \eqref{eq:assumption upper} can be interpreted, summing the two relations in \eqref{trans+} and \eqref{trans-}, as the total intensity of the process $A(t)$ (or, equivalently, of $X(t)$).}

As special case of Theorem \ref{teo:main upper} for constant control policies, we have the following result.

\begin{corollary}\label{teo:constant} Consider a controlled evolutionary dynamics $(\mc G, \beta, \u)$ under constant control policy $U(t)=\u$  {with initial configuration $X(0)=\0$} and let $\phi$ be the minimum conductance profile of $\mc G$, { defined in \eqref{eq:min cond}}. Then, the conditional expected spreading time satisfies
\be\label{perf-guarantee}
\E_{\0}[T]\leq\frac{\beta}{(2\beta-1)\1^\top \u}+ \frac{1}{2\beta-1}\sum_{a=1}^{n-1}\frac{\text{1}}{\phi(a)}.
\ee
\end{corollary}
\begin{IEEEproof}
Observe that, for every $a\in\{1,\dots,n\}$, for all subsets $S\subset \mc V$ with $\abs{S}=a$, we have $\phi(a)\leq{{\zeta(S)}}$. Therefore, if $A(t)=a$, then $B(t)+C(t)\geq\phi(a)$. If $A(t)=0$, then $C(t)=\1^\top \u$. Finally,~\eqref{bound F} from Theorem~\ref{teo:main upper} is applied with $f(a)=\phi(a)$, for $a\neq 0$, and $f(0)=\1^\top \u$.
\end{IEEEproof}


\subsection{Fundamental Performance Limitation}\label{sec:lower}

In this subsection we present a series of results illustrating fundamental limitations of the controlled evolutionary {dynamics}. We start by establishing two preliminary monotonicity properties of the process, which will be instrumental to our main results. For the sake of readability, proofs are reported in Appendices~\ref{app:lemma1} and~\ref{app:lemma2}, respectively.

\begin{lemma}\label{lemma:monotonicity} Let $(\mc G,\beta, U(t))$ and $(\mc G,\gamma, U(t))$ be two controlled evolutionary dynamics and let  $X(t)$ and $Y(t)$ be, respectively, the corresponding Markov processes with initial conditions $X(0)=\x_0$ and $Y(0)=\y_0$.
If $\beta\leq\gamma$ and $\x_0\le \y_0$, then {$\E_{\y_0}[T_Y]\leq\E_{\x_0}[T_X]$ and $\E_{\y_0}[J_Y]\leq\E_{\x_0}[J_X]$}, where the subscripts of $T$ and $J$ refer to the processes they are associated to.
\end{lemma}\smallskip

\begin{lemma}\label{lemma:monotonicity2} Let $(\mc G,1, U(t))$ and $(\mc G,1,0)$ be two controlled evolutionary dynamics and let  $X(t)$ and $Y(t)$ be, respectively, the corresponding Markov processes with initial conditions $X(0)=\x_0$ and $Y(0)=\y_0$. If $\x_0\leq \y_0$ and $(\1-\y_0)^\top U(t)=0$, for every $t\ge0$, i.e., if the {uncontrolled process $Y(t)$ has initial condition equal to $1$ in all the nodes in which the control is exerted in the controlled process $X(t)$, then, $\E_{\y_0}[T_Y]\leq\E_{\x_0}[T_X]$}, where the subscripts of $T$ and $J$ refer to the processes they are associated to.
\end{lemma}\smallskip

{
\begin{remark}\label{rem:different}  These results have straightforward consequences:
\begin{enumerate}
\item Lemma \ref{lemma:monotonicity} implies that any lower bound on the expected spreading time obtained for the case when $\beta=1$ will automatically yield a lower bound for every value of $\beta$. 
\item Lemma~\ref{lemma:monotonicity2} implies that, for a controlled evolutionary dynamics where the novel state always wins (i.e., when $\beta=1$), {it is always possible to establish a lower bound on the expected spreading time by considering an uncontrolled process with initial condition equal to $1$ in all the nodes in which the external control is exerted.} We wish to emphasize that controlled evolutionary dynamics with $\beta<1$ do not, {in general,} enjoy this monotonicity property.
\end{enumerate}
\end{remark}\smallskip

The previous results motivate a deeper analysis of  the controlled evolutionary dynamics of type $(\mc G,1, U(t))$. In this case, if started with an initial configuration $X(0)$ such that $\1^\top X(0)=k$, the Markov process $X(t)$,  will always undergo exactly $n-k$ transitions before being absorbed in the all-$1$ configuration. {Indeed, for $\beta=1$~\eqref{eq:transitions} shows that $\lambda_i^-(\x,t)=0$ for every configuration $\x$, node $i$ and time $t\geq 0$. This says that the process can only undergo transitions where the number of $1$'s is increasing, either driven by the spreading mechanism or by the external control.}  

{We let $0=T_k<T_{k+1}<\cdots <T_{n}=T$ be the random times at which these $n-k$ jumps occur and we denote by $X_h=X(T_h^+)$ the configurations of the process after the corresponding jumps, so that $X_k=X(0)$ and $X_n=\1$. According to this notation,  $T_h$ is the time of the jump from $X_{h-1}$ to $X_h$. Finally, we let $B_h$ to be the corresponding values for the boundary of the process when in configuration $X_h$, that is  $B_h=\zeta(X_h)$.} This time process can be described recursively as follows. We first let $\sigma_h$ to be the $\sigma$-algebra generated by $T_h$ and by the process $X(t)$ for $t\leq T_{h}$. Given $\sigma_{h-1}$ we consider two independent random variables $t_h^{\text{s}}$ and $t_h^{\text{c}}$ whose distribution functions are given, respectively, by
\be\label{times}\ba{lll}\P[t_{h}^{\text{s}}\geq t\,|\, \sigma_{h-1}]&=&\exp\left({-B_{h-1}t}\right),\\\P[t_{h}^{\text{c}}\geq t\,|\, \sigma_{h-1}]&=&\ds\exp\left({-\int_{T_{h-1}}^{T_{h-1}+t}C(s)\text{d}s}\right).\ea\ee
We then put $t_h=\min\{t_{h}^{\text{s}}, t_{h}^{\text{c}}\}$. The interpretation of $t_h$ is the following: from state $X_{h-1}$ the system can evolve either through the spreading mechanisms or through an occurrence of the external control in a node currently in state $0$; $t_{h}^{\text{s}}$ and $t_{h}^{\text{c}}$ are exponentially distributed {random} variables that model the random times at which the two phenomena would independently take place, according to their definition. {According to the properties of Markov processes~\cite{Levin2009}, the minimum of these two random variables models the time to wait for the next jump, that is, $T_h=T_{h-1}+t_h$.} Define
\rev{$$ c_h=\int_{T_{h-1}}^{T_{h}}C(s)\text{d}s$$
to be the effective control rate exerted during the interval $[T_{h-1}, T_h]$. The following result shows that there is an exact algebraic relation that ties the average length of the interval $[T_{h-1}, T_h]$, the average effective control rate in this interval $c_h$, and the active boundary after the $(h-1)$-th jump $B_{h-1}$.
\begin{prop}\label{prop:fund-limit} For every $h$ such that $B_{h-1}\neq 0$, the following relation holds true
\rev{\be\label{fund-relation}\E[T_h-T_{h-1}\,|\, \sigma_{h-1}]=\ds\frac{1-\E[c_h\,|\, \sigma_{h-1}]}{B_{h-1}}\,.\ee}
\end{prop}}

\begin{IEEEproof} 
First, observe that $T_h-T_{h-1}=t_h=\min\{t_{h}^{\text{s}}, t_{h}^{\text{c}}\}\,.$ Using the property of the minimum of two independent exponentially distributed random variables, we note that
\be\label{t-ripartition}\P[t_{h}\geq t\,|\, \sigma_{h-1}]=\ds\exp\left({-\int_{T_{h-1}}^{T_{h-1}+t}\big(B_{h-1}+C(s)\big)\text{d}s}\right),\ee
which, in particular, yields
$$\frac{d}{dt}\P[t_{h}\geq t\,|\, \sigma_{h-1}]=\big(B_{h-1}+C(T_{h-1}+t)\big)\P[t_{h}\geq t\,|\, \sigma_{h-1}],$$
From this and the fact that $T_{h-1}$ and $\{C(s)\,|\, s\in [T_{h-1}, T_h]\}$ are both $\sigma_{h-1}$-measurable, we have 
\be\label{fund-relation-proof}\ba{rcl}\E[c_h\,|\, \sigma_{h-1}] 
&=& \E\left[\ds\int\limits_{T_{h-1}}^{+\infty}C(s){{\mathbbm 1}}_{[T_{h-1}, T_h]}(s)\text{d}s\,|\,\sigma_{h-1}\right]\\
&=&\ds\int\limits_{T_{h-1}}^{+\infty}C(s) \E\left[{{\mathbbm 1}}_{[T_{h-1}, T_h](s)}\,|\,\sigma_{h-1}\right]\text{d}s\\
&=&\ds\int\limits_{T_{h-1}}^{+\infty}C(s) \P[t_h\geq s-T_{h-1}\,|\,\sigma_{h-1}]\text{d}s\\
&=&\ds\int\limits_{T_{h-1}}^{+\infty}-\frac{d}{ds} \P[t_h\geq s-T_{h-1}\,|\,\sigma_{h-1}]\text{d}s\\
&-& B_{h-1}\ds\int\limits_{T_{h-1}}^{+\infty} \P[t_h\geq s-T_{h-1}\,|\,\sigma_{h-1}]\text{d}s\\
&=& 1-B_{h-1}\E[t_h,|\, \sigma_{h-1}] ,\ea\ee
which yields the proof.
\end{IEEEproof}\medskip

{If we further define
\be\label{J_h}J_h:=\int\limits_{T_{h-1}}^{T_{h}}\1^\top U(s)\text{d}s\ee
to be the control rate exerted during the interval $[T_{h-1}, T_h]$, we immediately observe that, by definition, $J_h\geq c_h$.} \eqref{fund-relation} yields the following result.

\begin{corollary}\label{cor:fund} Assumethat the initial condition $X(0)=\x_0$ is such that $\1^\top \x_0=k>0$. Then,
\be\label{fund-lim} \ba{lll}\E_{\x_0}[T] 
&\geq&\ds\sum\limits_{h=k+1}^n\E\left[\ds\frac{1-\E[J_h\,|\, \sigma_{h-1}]}{B_{h-1}}\right],\\[8pt]
\E_{\x_0}[J] &=&\ds\sum\limits_{h=k+1}^n\E\big[\E[J_h\,|\, \sigma_{h-1}]\big]\,.\ea
\ee 
\end{corollary}
\begin{IEEEproof} 
The former follows by summing and averaging \eqref{fund-relation}, using that {$B_h\neq 0$ for $h=k,\dots , n-1$}; the latter is the definition of $J$, {obtained by summing all the terms in \eqref{J_h}.}
\end{IEEEproof}

\begin{remark}
In the case when $X(0)=\0$, the relations in \eqref{fund-lim} have to be modified as follows. The first jump is necessarily triggered by an exogenous activation, which implies
\be\label{eq:t1_0}\E[t_1]=\int\limits_0^{+\infty}\exp\left({-\int\limits_{0}^{t}\1^\top U(s)\text{d}s}\right)\text{d}s.\ee
Moreover, \eqref{fund-relation-proof} yields $\E[c_1]=\E[J_1]=1$. Therefore, \eqref{fund-lim} is modified by setting $k=1$ and adding the term $\E[t_1]$ from \eqref{eq:t1_0} to the {right-hand side of the inequality for $\E_{\0}[T]$} and $1$ to {the right-hand side of the equality for $\E_{\0}[J]$.}
\end{remark}\smallskip

The applicability of \eqref{fund-lim}, in the present general form, is limited by the fact that the control efforts and the boundary evolutions can not be uncoupled when the averaging operation is taken. For {$\mc U$-controlled evolutionary dynamics}, Corollary \ref{cor:fund} can be relaxed, yielding more explicit, even if less tight, bounds that turn to be useful in case when $\mc U$ is sufficiently small.

\begin{corollary}\label{cor:fund-limit} Let $(\mc G,\beta, U(t))$ be a $\mc U$-controlled evolutionary dynamics with initial condition $X(0)=\x_0\leq\delta^{(\mc U)}$. Then,
\begin{equation}\label{eq:minimization time2}
{\E_{\x_0}[T]}\geq \sum_{h=|\mc U|}^{n-1}\frac{1}{\eta(h)}\,,
\end{equation}
where $\eta$ is defined as in~\eqref{eq:max exp}.
\end{corollary}
\begin{IEEEproof}
It follows from Lemmas \ref{lemma:monotonicity} and  \ref{lemma:monotonicity2} that it is sufficient to prove the bound for the controlled evolutionary dynamics $(\mc G,1, 0)$ with initial condition $Y(0)=\delta^{(\mc U)}$. In this case, \eqref{fund-lim} {reduces to}
$$\E_{\x_0}[T]\geq \E_{\delta^{(\mc U)}}[T_Y]\geq \sum\limits_{h=|\mc U|}^{n-1}\E\left[\ds\frac{1}{B_{h}}\right].
$$
Finally, the claim follows from the definition of $\eta(h)$ in~\eqref{eq:max exp}.
\end{IEEEproof}\smallskip

For constant control policies with novel-free initial condition, we can {refine the result as follows.}

\begin{corollary}\label{cor:fund-limit2} Let $(\mc G,\beta, \u)$ be a constant $\mc U$-controlled evolutionary dynamics with initial condition $X(0)=\0$. Then, the expected spreading time verifies
\begin{equation}\label{eq:minimization time3}
\E_{\0}[T]\geq \frac{1}{\1^\top \u}+\sum_{h=|\mc U|}^{n-1}\frac{1}{\eta(h)}.
\end{equation}
\end{corollary}
\begin{IEEEproof}
{Consider the {random time  $t_1$ corresponding to} the first jump of the process. From \eqref{eq:t1_0}, using that $\1^\top \u$ is constant, we compute $\E[t_1]=1/\1^\top \u$} and applying Corollary \ref{cor:fund-limit} with initial condition $X(t_1)\leq\delta^{(\mc U)}$ we obtain the claim.
\end{IEEEproof}\smallskip

By comparing the lower bound in \eqref{eq:minimization time3} with the {corresponding} upper bound in \eqref{perf-guarantee}, we conclude that, for constant $\mc U$-controlled evolutionary dynamics,
the control rate $\1^\top \u$ has a limited effect on the performance. In particular, $\E_{\0}[T]$ remains bounded away from $0$ even in the limit case as $\1^\top \u\to +\infty$, even though the expected control cost $\E_{\0}[J]=(\1^\top \u)\E_{\0}[T]$ grows large. 

Equations \eqref{eq:minimization time2} and \eqref{eq:minimization time3} moreover suggest that{, together with the network structure,} the support $\mc U$ of the control action may play a key role in achieving suitable spreading performance. In this direction, we propose another estimation of the expected spreading time {in which $\mc U$ has a central role}.

\begin{corollary}\label{cor:fund-limit3} Let $(\mc G,\beta, U(t))$ be a $\mc U$-controlled evolutionary dynamics with initial condition $X(0)=\0$. Then,
$$
\E_{\0}[T]\geq \left(\min\limits_{\mc R\supseteq\mc U}{{\zeta(\mc R)}}\right)^{-1}.
$$
\end{corollary}
\begin{IEEEproof}
{Using Lemmas \ref{lemma:monotonicity} and  \ref{lemma:monotonicity2}, we bound the expected spreading time by considering a controlled evolutionary dynamics $(\mc G,1, 0)$ with initial condition $Y(0)=\delta^{(\mc R)}$ where $\mc R\supseteq\mc U$.} \eqref{fund-lim} yields
$$\E_{\0}[T]\geq \E_{\delta^{(\mc R)}}[T_Y]\geq \sum\limits_{h=|\mc R|}^{n-1}\E\left[\ds\frac{1}{B_{h}}\right]\geq 
\frac{1}{{\zeta(\mc R)}},
$$
where the last equality follows from the fact that $B_{|\mc R|}$ is the boundary at the initial condition $Y(0)=\delta^{(\mc R)}$. 
Since the above inequality holds true for every $\mc R\supseteq \mc U$, we obtain the claim.
\end{IEEEproof}\smallskip

We conclude our analysis with a useful equivalent description of the control cost term. The following relation expresses the term $\E[c_h\,|\, \sigma_{h-1}]$ as the probability that the $h$th jump in the process is due to an exogenous event.

\begin{prop}\label{prop:cost-equiv} The following relation holds true
$$\E[c_h\,|\, \sigma_{h-1}]=\P[t_{h}^{\text{c}}<t_{h}^{\text{s}}\,|\, \sigma_{h-1}].$$
\end{prop}
\begin{IEEEproof}
It follows from \eqref{times} that
$$\ba{l}\P[t_{h}^{\text{c}}<t_{h}^{\text{s}}\,|\, \sigma_{h-1}]=\P[t_{h}^{\text{s}}-t_{h}^{\text{c}}>0\,|\, \sigma_{h-1}]
\\=\ds\int\limits_{0}^{+\infty}B_{h-1}e^{-B_{h-1}t}\left[1-\exp\left({-\ds\int_{T_{h-1}}^{T_{h-1}+t}C(s)\text{d}s}\right)\right]\text{d}t\\
=1-B_{h-1}\ds\exp\left({-\int_{T_{h-1}}^{T_{h-1}+t}\big(B_{h-1}+C(s)\big)\text{d}s}\right).
\ea
$$
The claim then follows using \eqref{t-ripartition} and \eqref{fund-relation-proof}.
\end{IEEEproof}\smallskip

Denote now by $N^{\text{c}}$ the random variable counting the total number of activations due to exerting the control action, more formally in the notation above:
\be\label{nc} N^{\text{c}}=\left|\{h=1,\dots , n: t_{h}^{\text{c}}<t_{h}^{\text{s}}\}\right|.\ee
Summing over $h$ and averaging the second relation in Proposition \ref{prop:cost-equiv}, we obtain that 
\be\label{control-cost}\E[N^{\text{c}}]\leq \E[J].\ee 
In other terms, the expected number of activations due to the control action is a lower bound on the control cost.}

\section{Three fundamental examples}\label{sec:constant} 
In this section, we formally introduce and analyze three key examples of graph families parameterized by their size $n$: {expander graphs, stochastic block models (SBM), and ring graphs}.  For each of them, we fully analyze the behavior of the corresponding controlled evolutionary {dynamics} and discuss the corresponding performance behavior as a function of $n$. 

\rev{Expander graphs and ring graphs constitute extreme cases of highly and poorly connected structures, respectively, and, as we will see, they represent benchmarks for topologies that are easy to control and hard to control, respectively. SBMs, instead, are representative of real-world geographical networks~\cite{Fortunato2016} and display an intermediate behavior: no fundamental limitation precludes fast diffusion, but constant control policies may fail in achieving it. In Section \ref{sec:feedback} we show that, for SBMs,  feedback control policies play a crucial role.}

For the sake of simplicity, in this section we consider graphs $\mc G$ {in which all the links in $\mathcal E$ have the same weight, that is,} $W_{ij}=w$, for all  $\{i,j\}\in\mc E$. Also, we assume that the value of $w$ is scaled with respect to the maximum degree $\Delta$ of the nodes in the graph, namely, we take $w=\alpha/\Delta$ where $\alpha>0$ is a constant kept fixed as $n$ varies. {Such a choice prevents the sum of the weights insisting on each node from blowing up (which would be inconsistent with any real-world application) and guarantees a fair comparison between the topologies in the three examples.}

We consider controlled evolutionary dynamics on $\mc G$, $(\mc G,\beta, U(t))$ always with initial condition $X(0)=\0$.
First, we notice that the trivial bound $\eta(h)\leq \alpha h$ substituted in \eqref{eq:minimization time2} yields, for every $\mc U$-controlled evolutionary dynamics, {the following bound on the expected spreading time}
\begin{equation}\label{log-time}
\E_{\0}[T]\geq \sum_{h=|\mc U|}^{n-1}\frac{1}{\alpha h}\geq \frac{1}{\alpha}\log\frac{n}{|\mc U|}.
\end{equation}
Hence, if $\mc U$ is assumed to be constant in $n$, the expected spreading time grows at least logarithmically in $n$.

\begin{figure}
\begin{center}
\subfloat[Complete graph]{	
\begin {tikzpicture}
\def \n {9}
\def \radius {1cm}
\def \margin {10} 

\foreach \s in {1,...,\n}
{
  \node[draw, circle] (\s) at ({360/\n * (\s - 1)}:\radius) {};
}
\foreach \s in {1,...,\n}
{
\foreach \t in {1,...,\n}
{
\draw[-, >=latex] (\s) edge (\t);
}
}

\end{tikzpicture}}\,\subfloat[Stochastic block model]{	
\begin {tikzpicture}
\def \n {6};
\def \m {5};
\def \r {0.65};
\def \phi {20};
\foreach \t in {1,...,\n}
{
\node[draw, circle] (\t) at ({\r*sin(360*\t/\n)},{\r*cos(360*\t/\n)}) {};
}

\foreach \t in {1,...,\m}
{
\node[draw, circle] (1\t) at ({1.8+\r*sin(360*\t/\m+\phi)},{.72+\r*cos(360*\t/\m+\phi)}) {};
}

\foreach \s/\t in {1/2, 1/4, 1/5,2/3, 2/5, 3/4,3/6,5/6, 11/13,11/14,12/15,12/13,14/15,11/15,1/13,2/12,12/14,12/11}
{
\draw[-, >=latex] (\s) edge (\t);
}

\end{tikzpicture}}\label{fig:topologies:b}\,\subfloat[Ring graph]{	
\begin {tikzpicture}
\def \n {9}
\def \radius {1cm}
\def \margin {10} 

\foreach \s in {1,...,\n}
{
  \node[draw, circle] (\s) at ({360/\n * (\s - 1)}:\radius) {};
}
\foreach \s/\t in {1/2,2/3,3/4,4/5,5/6,5/6,6/7,7/8,8/9,9/1}
{
\draw[-, >=latex] (\s) edge (\t);
}

\end{tikzpicture}}
\end{center}
\caption{Examples of the topologies analyzed in this paper.}
\label{fig:topologies}
\end{figure}
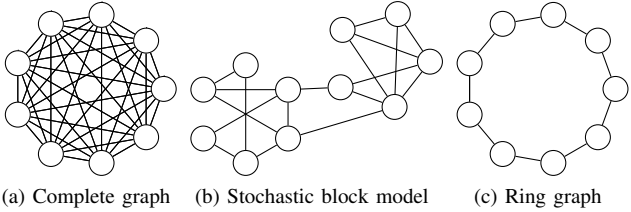

\subsection{Expander graphs}
A family of graphs is said to be an \emph{expander} if $\exists\,\gamma>0$ such that the minimum conductance profile verifies $\phi(h)\geq\gamma\min\{h,n-h\}$, for all $n$ and $h=1,\dots,n-1$. {Expander graphs are often used to represent well connected real-world systems. Classical models of small-world and scale-free networks, often adopted to model social networks, belong to this family.} 

We now show how these networks exhibit fast spread. Precisely we demonstrate that, using constant control policies with bounded nonzero rate, the expected spreading time scales logarithmically with respect to the network size.

Let $\mc G$ be an expander graph with $n$ nodes.  
Using the expander condition  on $\phi(h)$ in Corollary~\ref{teo:constant}, we obtain
\be \label{eq:complete_time}
\E_{\0}[T]\leq\frac{\beta}{(2\beta-1)\1^\top \u}+\frac{2\ln(n/2)+2}{\gamma(2\beta-1)}.
\ee
If we stick to control policies supported on a fixed subset $\mc U$, the lower bound in \eqref{log-time} concludes that expander graphs are a benchmark for fast-diffusive topologies. Figure~\ref{fig:expander} shows Monte Carlo estimations of the expected spreading time for complete graphs (in Figure \ref{fig:topologies}(a)), together with our analytical bounds. {Note that the bound in \eqref{eq:complete_time} is tight for the complete graph. {In fact, for complete graphs the inequality~\eqref{eq:assumption upper} holds true as an equality when the function $f(h)$ coincides with the minimum conductance profile (since all the subsets with $h$ nodes have the same weighted boundary). This suggests that the spreading time is close to the upper bound \eqref{bound F} in Theorem~\ref{teo:main upper} when one can determine a function $f(h)$ so that the inequality~\eqref{eq:assumption upper} is close to an equality.}
}
\begin{figure}
\centering
\subfloat[$\beta=0.7$]{\definecolor{mycolor1}{rgb}{0.00000,0.44700,0.74100}%
\begin{tikzpicture}

\begin{axis}[%
 axis lines=middle,
    axis line style={->},width=\l  cm,
height=\h cm,
at={(0.741in,0.457in)},
scale only axis,
xmin=00,
xmax=1100,
xlabel={$n$},
 extra y ticks ={0},
    extra y tick labels={$0$},
     extra x ticks ={0},
    extra x tick labels={$0$},
ymin=0,
ymax=44,
ylabel={$\E_{\0}[T]$},
axis background/.style={fill=white},
]
\addplot [only marks, mark size=0.9, color=mycolor1,solid,forget plot]   
 plot [error bars/.cd, y dir = both, y explicit]
 table[row sep=crcr, y error plus index=2, y error minus index=3]{%
50	20.5629083989854	3.35210041455279	3.35210041455279\\
100	24.3244563297278	3.21053373912277	3.21053373912277\\
150	25.4409197404178	3.45783657701288	3.45783657701288\\
200	28.0111085421568	3.66531197123721	3.66531197123721\\
250	29.2785501951727	3.9124234752146	3.9124234752146\\
300	29.2378030578623	3.31485533890742	3.31485533890742\\
350	30.2150718290018	3.84807428784269	3.84807428784269\\
400	30.6972400129549	3.56672720300437	3.56672720300437\\
450	30.9862498257935	3.5310346604099	3.5310346604099\\
500	32.7835792744218	3.94909349520207	3.94909349520207\\
550	31.8044017845526	3.23869060586757	3.23869060586757\\
600	32.9509509138425	3.42039193760885	3.42039193760885\\
650	33.7188193540301	2.99130426235215	2.99130426235215\\
700	33.8198853343954	3.44354374471879	3.44354374471879\\
750	34.2455009289105	3.32360650597251	3.32360650597251\\
800	34.7157594794453	3.37239921763916	3.37239921763916\\
850	35.0093415805252	3.33275175025922	3.33275175025922\\
900	35.2953622572879	3.94127854549157	3.94127854549157\\
950	35.6806653764764	3.83759976462214	3.83759976462214\\
1000	35.1165382832664	3.54175078999668	3.54175078999668\\
};

\addplot [color=mycolor1,solid,forget plot,thick]
  table[row sep=crcr]{%
20	18.2629254649702\\
40	21.72866136777\\
60	23.7559869083108\\
80	25.1943972705697\\
100	26.3101150271407\\
120	27.2217228111105\\
140	27.9924762102468\\
160	28.6601331733694\\
180	29.2490483516513\\
200	29.7758509299405\\
220	30.2524018289621\\
240	30.6874587139102\\
260	31.0876722522779\\
280	31.4582121130465\\
300	31.8031764704813\\
320	32.1258690761691\\
340	32.4289921852513\\
360	32.7147842544511\\
380	32.9851203608024\\
400	33.2415868327402\\
420	33.4855376535874\\
440	33.7181377317618\\
460	33.940396544616\\
480	34.15319461671\\
500	34.3573045893112\\
520	34.5534081550776\\
540	34.7421097949919\\
560	34.9239480158463\\
580	35.0994046149026\\
600	35.268912373281\\
620	35.432861487396\\
640	35.5916049789689\\
660	35.7454632723026\\
680	35.894728088051\\
700	36.0396657724173\\
720	36.1805201572508\\
740	36.3175150281914\\
760	36.4508562636022\\
780	36.5807336956185\\
800	36.7073227355399\\
820	36.8307857984918\\
840	36.9512735563871\\
860	37.068926043438\\
880	37.1838736345615\\
900	37.2962379138218\\
920	37.4061324474157\\
940	37.5136634735205\\
960	37.6189305195097\\
980	37.7220269555234\\
1000	37.823040492111\\
};
\addplot [color=mycolor1,solid,forget plot,thick]
  table[row sep=crcr]{%
20	2.99573227355399\\
40	3.68887945411394\\
60	4.0943445622221\\
80	4.38202663467388\\
100	4.60517018598809\\
120	4.78749174278205\\
140	4.9416424226093\\
160	5.07517381523383\\
180	5.19295685089021\\
200	5.29831736654804\\
220	5.39362754635236\\
240	5.48063892334199\\
260	5.56068163101553\\
280	5.63478960316925\\
300	5.7037824746562\\
320	5.76832099579377\\
340	5.82894561761021\\
360	5.88610403145016\\
380	5.94017125272043\\
400	5.99146454710798\\
420	6.04025471127741\\
440	6.08677472691231\\
460	6.13122648948314\\
480	6.17378610390194\\
500	6.21460809842219\\
520	6.25382881157547\\
540	6.29156913955832\\
560	6.32793678372919\\
580	6.36302810354047\\
600	6.39692965521615\\
620	6.42971947803914\\
640	6.46146817635372\\
660	6.49223983502047\\
680	6.52209279817015\\
700	6.5510803350434\\
720	6.5792512120101\\
740	6.60665018619822\\
760	6.63331843328038\\
780	6.65929391968364\\
800	6.68461172766793\\
820	6.7093043402583\\
840	6.73340189183736\\
860	6.75693238924755\\
880	6.77992190747225\\
900	6.80239476332431\\
920	6.82437367004309\\
940	6.84587987526405\\
960	6.86693328446188\\
980	6.88755257166462\\
1000	6.90775527898214\\
};
\end{axis}
\end{tikzpicture}%
}\,\,\subfloat[$\beta=0.8$]{\begin{tikzpicture}

\begin{axis}[%
 axis lines=middle,
    axis line style={->},width=\l  cm,
height=\h cm,
at={(0.741in,0.457in)},
scale only axis,
xmin=00,
xmax=1100,
xlabel={$n$},
ymin=0,
ymax=34,
ylabel={$\E_{\0}[T]$},
axis background/.style={fill=white},
 extra y ticks ={0},
    extra y tick labels={$0$},
     extra x ticks ={0},
    extra x tick labels={$0$},
]
\addplot [only marks, mark size=0.9, color=red,solid,forget plot] 
 plot [error bars/.cd, y dir = both, y explicit]
 table[row sep=crcr, y error plus index=2, y error minus index=3]{%
50	14.0390786480224	3.35078277928833	3.35078277928833\\
100	16.2348632131287	3.86217211923306	3.86217211923306\\
150	17.9223226162453	3.95165889176265	3.95165889176265\\
200	19.0405910209865	3.83023956801408	3.83023956801408\\
250	19.4058361551386	3.74963143552334	3.74963143552334\\
300	20.1764651445273	3.76923301772382	3.76923301772382\\
350	20.9970705582638	3.4562526282857	3.4562526282857\\
400	21.2654854655475	3.66406148198681	3.66406148198681\\
450	21.4636294198978	3.69802300731857	3.69802300731857\\
500	22.1259691392993	3.90449149438569	3.90449149438569\\
550	22.6103127303778	3.63588487481252	3.63588487481252\\
600	23.0856048461891	3.31147742646699	3.31147742646699\\
650	22.8002882997755	3.29043772838054	3.29043772838054\\
700	23.421523068026	3.87864289411554	3.87864289411554\\
750	23.3036196650322	3.88843040905517	3.88843040905517\\
800	23.3076614734744	3.27844162138958	3.27844162138958\\
850	24.1182389938668	3.12393332355427	3.12393332355427\\
900	23.8576896761272	3.42291729468391	3.42291729468391\\
950	24.1278772987637	3.39528887195951	3.39528887195951\\
1000	24.512191009791	3.59567097069415	3.59567097069415\\
};

\addplot [color=red,solid,forget plot,thick]
  table[row sep=crcr]{%
20	2.99573227355399\\
40	3.68887945411394\\
60	4.0943445622221\\
80	4.38202663467388\\
100	4.60517018598809\\
120	4.78749174278205\\
140	4.9416424226093\\
160	5.07517381523383\\
180	5.19295685089021\\
200	5.29831736654804\\
220	5.39362754635236\\
240	5.48063892334199\\
260	5.56068163101553\\
280	5.63478960316925\\
300	5.7037824746562\\
320	5.76832099579377\\
340	5.82894561761021\\
360	5.88610403145016\\
380	5.94017125272043\\
400	5.99146454710798\\
420	6.04025471127741\\
440	6.08677472691231\\
460	6.13122648948314\\
480	6.17378610390194\\
500	6.21460809842219\\
520	6.25382881157547\\
540	6.29156913955832\\
560	6.32793678372919\\
580	6.36302810354047\\
600	6.39692965521615\\
620	6.42971947803914\\
640	6.46146817635372\\
660	6.49223983502047\\
680	6.52209279817015\\
700	6.5510803350434\\
720	6.5792512120101\\
740	6.60665018619822\\
760	6.63331843328038\\
780	6.65929391968364\\
800	6.68461172766793\\
820	6.7093043402583\\
840	6.73340189183736\\
860	6.75693238924755\\
880	6.77992190747225\\
900	6.80239476332431\\
920	6.82437367004309\\
940	6.84587987526405\\
960	6.86693328446188\\
980	6.88755257166462\\
1000	6.90775527898214\\
};

\addplot [color=red,solid,forget plot,thick]
  table[row sep=crcr]{%
20	12.3419503099802\\
40	14.6524409118466\\
60	16.0039912722072\\
80	16.9629315137131\\
100	17.7067433514272\\
120	18.3144818740737\\
140	18.8283174734979\\
160	19.2734221155796\\
180	19.6660322344342\\
200	20.0172339532936\\
220	20.3349345526414\\
240	20.6249724759401\\
260	20.8917815015186\\
280	21.1388080753643\\
300	21.3687843136542\\
320	21.5839127174461\\
340	21.7859947901675\\
360	21.9765228363007\\
380	22.1567469072016\\
400	22.3277245551601\\
420	22.4903584357249\\
440	22.6454251545079\\
460	22.7935976964106\\
480	22.9354630778066\\
500	23.0715363928741\\
520	23.2022721033851\\
540	23.3280731966612\\
560	23.4492986772308\\
580	23.5662697432684\\
600	23.6792749155207\\
620	23.7885743249306\\
640	23.8944033193126\\
660	23.9969755148684\\
680	24.096485392034\\
700	24.1931105149449\\
720	24.2870134381672\\
740	24.3783433521276\\
760	24.4672375090681\\
780	24.5538224637456\\
800	24.6382151570266\\
820	24.7205238656612\\
840	24.8008490375914\\
860	24.8792840289587\\
880	24.9559157563743\\
900	25.0308252758812\\
920	25.1040882982771\\
940	25.1757756490137\\
960	25.2459536796731\\
980	25.3146846370156\\
1000	25.3820269947406\\
};
\end{axis}
\end{tikzpicture}%
}
\caption{Monte Carlo estimation ($200$ simulations) of the expected spreading time $\E_{\0}[T]$ on complete graphs for different values of $n$ ($\alpha=1$), with $90\%$ confidence intervals. The two solid \rev{curves} are the theoretical bounds from \eqref{log-time} and \eqref{eq:complete_time}.}
\label{fig:expander}
\end{figure}
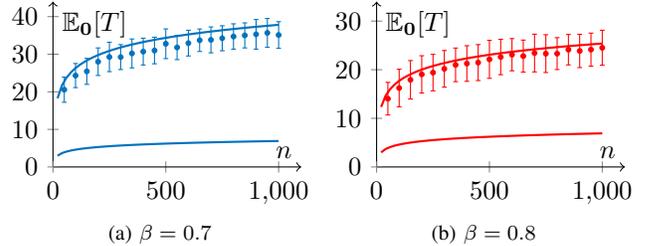

\subsection{Stochastic block models}\label{SBM}
As a second example we consider \emph{stochastic block models} (SBM), illustrated in Figure \ref{fig:topologies}(b). These graphs are composed of dense {(expander)} subgraphs linked among each other by few connections and have been often used in the literature to represent social, economical, and geographical structures~\cite{Holland1983,Barthelemy2011,Fortunato2016}. For the sake of simplicity, we limit our analysis to the case of two subgraphs (called communities), but our results can be easily generalized. Since SBMs are random graphs, results will be provided with high probability, i.e., with probability converging (at least polynomially) to $1$ as the network size $n$ grows. We will prove that spread is slow unless the control is exerted on all of its communities. 

{In this paper, we consider the following implementation of SBMs.} Fixed $c\in(0,1/2]$, we {partition the nodes into two disjoint sets $\mc V=\mc V^1\cup\mc V^2$, with $n_1=\lfloor cn\rfloor $ and $n_2=n- \lfloor cn\rfloor${nodes}, respectively. Without loss of generality, we assume $c\leq 1/2$. The nodes in each subset form a subgraph that is generated according to an Erd\H{o}s-R\'enyi random graph. In both subgraphs,}
 each link is present with probability $p\in(0,1]$, independent of the others~\cite{Bollobas2001}. On top of this, $L$ links positioned uniformly at random connect nodes belonging to the two different subgraphs. 

If the control is exerted in only one of the subgraphs, then the expected spreading time grows  linearly in $n$. This can be shown as follows. We fix $\mc U\subseteq \mc V^1$ and consider any $\mc U$-controlled evolutionary dynamics $(\mc G, \beta, U(t))$ on $\mc G$. Considering the standing assumptions that all non-zero weights are equal to $w=\alpha/\Delta$ and that the maximal degree satisfies the inequality $\Delta \geq n_2p\geq (1-c)np$ (w.h.p), we have that 
$
{\zeta(\mc V^1)}
\leq{\alpha L}/{(1-c)np}.$
Hence, Corollary \ref{cor:fund-limit3}, yields
\be\label{SBM-linear}\E_{\0}[T]\geq \frac{(1-c)np}{\alpha L}.\ee
{We demonstrate that such an estimation is asymptotically order-tight for constant control policies $U(t)=\u$ by leveraging Corollary \ref{teo:constant}.
%
%
To this aim, we estimate the minimum conductance profile as follows.} 
First, note that (even though SMBs are no expanders) each one of the two subgraphs is expander with $\gamma=cnpw/2$~\cite{Ganesh2005}. Using the trivial bound $\Delta \leq n$, we get $\gamma\geq c\alpha p/2$.
Given an integer $0<h<n$, any subset $\mc W \subseteq \mc V$ with $|\mc W|=h$ can be written as $\mc W=\mc W^1\cup\mc W^2$, with $\mc W^i\subseteq \mc V^i$, $|\mc W^i|=h_i$, and $h_1+h_2=h$. Therefore,
\be\ba{l}\label{eq:expansiveness}\ds\phi(h)\geq\frac{c\alpha p}{2}\cdot \\
\cdot\min\limits_{\tiny\ba{c}h_1,h_2\\h_1+h_2=h\ea}\left\{\min\{h_1,n_1-h_1\}+\ds\min\{h_2,n_2-h_2\}\right\}\\
=\ds\frac{c\alpha p}{2}\cdot\left\{\ba{ll}\min\{h, n_1-h\},\quad &h\leq n_1,\\ \min\{h-n_1, n_2-h\},\quad &n_1<h\leq n_2, \\ \min\{h-n_2, n-h\},\quad &n_2<h<n.
\ea\right.
\ea\ee
For $h\in\{n_1,n_2\}$, the previous bound {reduces to the trivial inequality $\phi(h)\geq0$}. 
However, the presence of $L$ links between the two subgraphs ensures that
\be\label{eq:inter}
\phi(h)\geq \frac{L\alpha}{n},\quad\text{if }h\in\{n_1,n_2\}.
\ee
Combining~\eqref{eq:expansiveness} and~\eqref{eq:inter}, using bounds on the harmonic series, and applying Corollary \ref{teo:constant}, we finally obtain
\be\label{SBM-linear2}\ba{lll}
\E_{\0}[T]&\leq&\ds\frac{1}{2\beta -1} \left(\frac{2}{L\alpha}n+\frac{12}{c\alpha p}\left(\ln\left(\frac n2\right)+1\right) \right)\\&&\ds+\frac{\beta}{(2\beta-1)\1^\top  \u}.\ea
\ee
This, together with \eqref{SBM-linear}, demonstrates that the expected spreading time {grows linearly with the network size $n$}. In Figure \ref{fig:sbm_fix} we show Monte Carlo estimations of the expected spreading time for a SBM (in Figure \ref{fig:topologies}(b)), together with our analytical bounds. 

Note that, if we allow the control nodes in both the constituent subgraphs of the SBM, we instead obtain a logarithmic growth lower bound on $\E_{\0}[T]$. \rev{These considerations extend to more general SBMs composed of many dense subgraphs. They lead to the conclusion that to obtain fast spreading time (of logarithmic growth) in these graphs using constant control policies or, more generally, control policies with constant support $\mc U$, the set $\mc U$ must necessarily have non-empty intersection with all the communities in the SBM. This is a drawback in practical applications as, not only it may require the cardinality of $\mc U$ to be large, but also needs precise a-priori information on the network topology to suitably position the control nodes.}



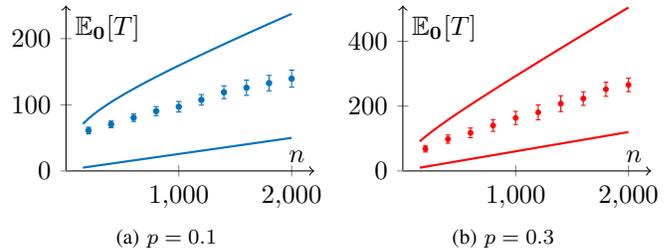
\begin{figure}
\centering
\subfloat[$p=0.1$]{\definecolor{mycolor1}{rgb}{0.00000,0.44700,0.74100}%
\begin{tikzpicture}
\begin{axis}[%
 axis lines=middle,
    axis line style={->},width=\l  cm,
height=\h cm,
at={(0.741in,0.457in)},
scale only axis,
xmin=0,
xmax=2200,
xlabel={$n$},
xtick={0,1000,2000},
ymin=0,
ymax=250,
ylabel={$\E_{\0}[T]$},
axis background/.style={fill=white},
 extra y ticks ={0},
    extra y tick labels={$0$},
]

\addplot [only marks, mark size=1, color=mycolor1,solid,forget plot] 
   plot [error bars/.cd, y dir = both, y explicit]
 table[row sep=crcr, y error plus index=2, y error minus index=2]{%
200	61.1469255513453	5.07174993626754	4507174993626754\\
400	70.4013968487594	5.10459062752116	5.10459062752116\\
600	80.4221188924956	5.64001666857414	5.64001666857414\\
800	90.4370773562378	6.7826441375093	6.7826441375093\\
1000	97.0888383673488	7.91475650056828	7.91475650056828\\
1200	107.524592282668	8.00656212490333	8.00656212490333\\
1400	119.039277070325	9.7197478034905	9.7197478034905\\
1600	125.826005190907	11.5274517664391	11.5274517664391\\
1800	132.838194773249	11.6098955192166	11.6098955192166\\
2000	139.557838351512	12.8896960347816	12.8896960347816\\
};

\addplot [color=mycolor1,solid,forget plot,thick]
  table[row sep=crcr]{%
150	71.0586489742685\\
200	79.1866835151315\\
250	86.239076037035\\
300	92.6111019836009\\
350	98.5136133140552\\
400	104.072469857797\\
450	109.368853785404\\
500	114.458195713034\\
550	119.380032043106\\
600	124.163554992933\\
650	128.830933454159\\
700	133.399399656721\\
750	137.882614181504\\
800	142.291589533796\\
850	146.635333230737\\
900	150.921306794736\\
950	155.155760482574\\
1000	159.3439820557\\
1050	163.490484791857\\
1100	167.599151719105\\
1150	171.673347761953\\
1200	175.716008002266\\
1250	179.729707910937\\
1300	183.716719796825\\
1350	187.679058596539\\
1400	191.61851933272\\
1450	195.536707996241\\
1500	199.435067190836\\
1550	203.314897571219\\
1600	207.177375876462\\
1650	211.023570187575\\
1700	214.854452906736\\
1750	218.670911854624\\
1800	222.473759804068\\
1850	226.263742707204\\
1900	230.04154682524\\
1950	233.807804931961\\
2000	237.563101731699\\
};
\addplot [color=mycolor1,solid,forget plot,thick]
  table[row sep=crcr]{%
150	5\\
2000	50\\
};

\end{axis}
\end{tikzpicture}%
}\,\,\subfloat[$p=0.3$]{\definecolor{mycolor1}{rgb}{0.00000,0.44700,0.74100}%
\begin{tikzpicture}
\begin{axis}[%
 axis lines=middle,
    axis line style={->},width=\l  cm,
height=\h cm,
at={(0.741in,0.457in)},
scale only axis,
xmin=0,
xmax=2200,
xlabel={$n$},
xtick={0,1000,2000},
ymin=0,
ymax=510,
ylabel={$\E_{\0}[T]$},
axis background/.style={fill=white},
 extra y ticks ={0},
    extra y tick labels={$0$},
]

\addplot [only marks, mark size=0.9, color=red,solid,forget plot] 
   plot [error bars/.cd, y dir = both, y explicit]
 table[row sep=crcr, y error plus index=2, y error minus index=2]{%
200	68.0234918432033	9.59516705253561	9.59516705253561\\
400	98.210098521369	12.7258485179532	12.7258485179532\\
600	117.476881147254	14.9681326028057	14.9681326028057\\
800	140.237487314037	18.1643351967695	18.1643351967695\\
1000	163.604516929381	20.2921138961486	20.2921138961486\\
1200	180.676820661436	22.9707927448385	22.9707927448385\\
1400	207.556514575639	24.5307971416595	24.5307971416595\\
1600	223.288046553103	20.9530134883363	20.9530134883363\\
1800	251.895640910938	21.8218585271794	21.8218585271794\\
2000	265.251489259813	20.6738371389374	20.6738371389374\\
};

\addplot [color=red,solid,forget plot,thick]
  table[row sep=crcr]{%
150	91.0586489742685\\
200	105.853350181798\\
250	119.572409370368\\
300	132.611101983601\\
350	145.180279980722\\
400	157.405803191131\\
450	169.368853785404\\
500	181.124862379701\\
550	192.71336537644\\
600	204.163554992933\\
650	215.497600120826\\
700	226.732732990054\\
750	237.882614181503\\
800	248.958256200463\\
850	259.96866656407\\
900	270.921306794736\\
950	281.822427149241\\
1000	292.677315389033\\
1050	303.490484791857\\
1100	314.265818385772\\
1150	325.006681095286\\
1200	335.716008002266\\
1250	346.396374577603\\
1300	357.050053130158\\
1350	367.679058596539\\
1400	378.285185999387\\
1450	388.870041329575\\
1500	399.435067190836\\
1550	409.981564237886\\
1600	420.510709209795\\
1650	431.023570187575\\
1700	441.521119573403\\
1750	452.004245187957\\
1800	462.473759804068\\
1850	472.93040937387\\
1900	483.374880158573\\
1950	493.807804931961\\
2000	504.229768398366\\
};
\addplot [color=red,solid,forget plot,thick]
  table[row sep=crcr]{%
150	10\\
2000	120\\
};

\end{axis}
\end{tikzpicture}%
}
\caption{Monte Carlo estimation ($200$ simulations) of the expected spreading time $\E_{\0}[T]$ on SBMs with $\u=\delta^{(1)}$, $\alpha=1$, $\beta=0.8$, $c=0.4$ and $L=5$ for different values of $n$, with $90\%$ confidence intervals. \rev{The two solid curves are theoretical bounds from \eqref{SBM-linear} and~\eqref{SBM-linear2}.}} 
\label{fig:sbm_fix}
\end{figure}

\subsection{Ring graphs}

\rev{Our last case study concerns ring graphs that represent an extreme scenario of poorly connected networks. We fix 
an undirected ring $\mc G$ with $n$ nodes and each link weighted $w=\alpha/2$, as represented in Figure \ref{fig:topologies}(c). and we consider a {controlled evolutionary dynamics} $(\mc G, \beta, U(t))$ with $X(0)=\0$. }

\rev{We bound the performance comparing our process $X(t)$ with the Markov process $Y(t)$ associated with the {controlled evolutionary dynamics}  $(\mc G, 1, U(t))$ with $Y(0)=\0$. Notice that
Lemma~\ref{lemma:monotonicity} ensures that the expected spreading time verifies $\E_{\0}[T_X]\geq\E_{\0}[J_Y]$ and the expected control cost verifies  $\E_{\0}[J_X]\geq\E_{\0}[J_Y]$, where the subscripts denote the process.}

\rev{Our aim is to bound the performance of $Y(t)$ applying Corollary \ref{cor:fund}. 
To this aim, we start with two simple remarks. First,  for every $\mc W\subseteq \mc V$, we have $\zeta [\mc W]\leq\alpha \abs{\mc W}$. Second, the contagion mechanism cannot increase the boundary $B(t)$: it decreases by $\alpha$, if the two neighbors of the node that changes its state are both state-$1$ nodes, or, otherwise, it remains the same. Moreover, considering that every jump due to exerting a control action can increase the boundary by $\alpha$, we have, for every $h$,  \be\label{bound-cycle}B_{h}\leq \alpha N^{\text{c}}.\ee 
where $N^{\text{c}}$ is the total number of control activations, formally defined in \eqref{nc}.
Consider now the event \rev{$E:=\{N^{\text{c}}\leq 2 \E[N^{\text{c}}]\}$. }}
From \eqref{bound-cycle} and Markov inequality we obtain
$$\ba{lll} \E\left[\ds\frac{1-\E[J_h\,|\, \sigma_{h-1}]}{B_{h-1}}\right]&\geq& \ds \E\left[\ds\frac{1-\E[J_h\,|\, \sigma_{h-1}]}{\alpha  N^{\text{c}}}{\mathbbm 1}_E\right]\\[10pt]
&\geq&\ds\frac{\E\left[1-\E[J_h\,|\, \sigma_{h-1}\right]}{2\alpha\E[N^{\text{c}}]}\mathbb P[E]\\[10pt]
&\geq&\ds \frac{1-\E\left[\E[J_h\,|\, \sigma_{h-1}]\right]}{4\alpha\E[N^{\text{c}}]}\,.
\ea
$$
From Corollary \ref{cor:fund}, using the above computation and~\eqref{control-cost}, we finally obtain
$$\E_{\0}[T_X]\geq\E_{\0}[T_Y]\geq \ds\frac{n-\E_{\0}[J_Y]}{4 \E_{\0}[J_Y]}\geq \ds\frac{1}{4\E_{\0}[J_X]}n-\frac14.$$
{This allows us to conclude that the spread is slow on large-scale ring graphs, unless adopting a control policy whose cost $\E_{\0}[J]$ blows up with the network size.}

\section{Feedback Control Policies}\label{sec:feedback}
{ Through the examples discussed in the previous section}, we have unveiled different behavior of the controlled evolutionary dynamics, depending on the structure of the underlying network topologies. Specifically, we have characterized \emph{easy-to-control} network topologies (e.g., expander graphs) where simple constant control policies {can} guarantee fast spread. On the other hand, we have identified  \emph{hard-to-control} network topologies (e.g., rings) where any control policy fails to achieve fast spread. There are networks, like the stochastic block models, that belong to neither of these {two} classes, for which the location (and possibly the number) of nodes where the control is exerted plays a crucial role in determining the performance. In such scenarios, the use of feedback control policies may {allow one to achieve fast spread with reasonable control efforts, while simple open-loop control policies fail.}

{Our  proposed feedback policies are based on the observation that, in the presence of an evolutionary advantage for the novel states (i.e., with $\beta>1/2)$, the evolutionary dynamics naturally induces a positive feedback loop enhancing the spreading process. 
However, the presence of bottlenecks in the graph topology (as in the example of SBMs) may slow down such a diffusion process. The objective of the proposed control policies is thus to enforce such a positive feedback whenever the process enters in a bottleneck. Technically, our policies rely on} the following two considerations: (i) it is of no use to exert control on state-$1$ nodes: this {yields} $C(t)=\1^\top U(t)$; and (ii) in order to apply Theorem \ref{teo:main upper}, the feedback control law $U(t)=\nu(X(t))$ must {enforce a relation like \eqref{eq:assumption upper}.} This suggests to consider feedback laws whose instantaneous rate $\nu(X(t))$ only depends on the configuration $X(t)$ through the two aggregate variables $A(t)$ and $B(t)$. Moreover, relation~\eqref{eq:assumption upper} suggests that a {valuable control policy should compensate for the boundary when this becomes too small, in order to maintain the overall intensity of the process $B(t)+C(t)$ always sufficiently large}.

On the basis of these considerations, we design a feedback control policy that is determined by two ingredients: a {\emph{target function}} $\iota:\{0,1\}^n\to \{1,\dots , n\}$ selecting the node in which the control has to be exerted (as a function of the current configuration); and a \emph{rate function} 
$\mu:\{0,\dots , n\}\times \R_{+}\to\R_{+}$ selecting the control rate (as a function of the number of $1$'s in the current configuration, that is, $A(t)$, and of the boundary $B(t)$).
We assume the {target} function to be any function $\iota:\{0,1\}^n\to \{1,\dots , n\}$ such that $x_{\iota(x)}=0$ (i.e., we always choose to control a node that is currently in state $0$). 
{The rate function is assumed to have the following form. 
We fix a parameter $K>0$ and we put}
\begin{equation}\label{control}\mu(a,b)=\left\{\begin{array}{ll}
K-b&\,\text{if }a<n\text{ and }b< K,\\0&\,\text{otherwise.}
\end{array}\right.\end{equation} 
Finally, the feedback control law $\nu(X(t))$ is given by:
\begin{equation}\label{feedback-control}
\nu_i(X(t))=\left\{\begin{array}{ll} \mu\big(A(t),B(t)\big)\;&\text{if } i=\iota\big(X(t)\big),\\0\;&\text{else}.\end{array}\right.
\end{equation}
Briefly, in the selected node $i=\iota(X(t))$ we exert  a control action with rate $\mu$ as in \eqref{control}, while in all other nodes no control is exerted.

In order to analyze the expected spreading time of this feedback control policy, it is useful to consider the following \emph{floor} version of the conductance profile.

\begin{dfn}
For $K\geq0$, the \emph{K-floor conductance profile} of an undirected weighted graph $\mc G=( \mc V,\mc E, W)$ of order $\abs{ \mc V}=n$ is the function $\phi_K:\{1,\dots,n-1\}\to\R_{+}$ defined as
$$
\phi_K(a):=\max\{\phi(a),K\}=\max\left\{\min_{\mc W\subset  \mc V,\abs{\mc W}=a}{{\zeta(\mc W)}},K\right\}.
$$
\end{dfn}

{From Theorem~\ref{teo:main upper} we establish} the following upper bounds on the expected spreading time and on the expected cost of the feedback control policy in~\eqref{feedback-control}.\medskip

\begin{corollary}\label{cor:feedback}
Consider the feedback controlled evolutionary dynamics $(\mc G, \beta, \nu)$ with initial condition $X(0)=\0$, where $\nu$ follows the control policy~\eqref{feedback-control}. Then, the expected spreading time verifies
\begin{equation}\label{time control}
\E_{\0}[T]\leq \frac{\beta}{(2\beta-1)K}+\frac{1}{2\beta-1}\sum_{a=1}^{n-1}\frac{1}{\phi_K(a)},
\end{equation}
and the expected control cost is bounded as
\begin{equation}\label{cost control}
\E_{\0}[J]\leq \ds\frac{\beta}{2\beta-1}+\frac{\left|\left\{1\leq a<n\,:\,\phi(a)<K\right\}\right|}{2\beta-1} .
\end{equation}
\end{corollary}
\begin{IEEEproof}
From \eqref{control} it follows that, for all $t$ such that $A(t)<n$, $C(t)=\max\{K-B(t),0\}$.
For such values of $t$ we thus have
$$
B(t)+C(t)=\max\{B(t),K\}\geq\phi_K(A(t)).
$$
The upper bound~\eqref{time control} follows from Theorem~\ref{teo:main upper}.

The estimation of the expected control cost is performed as follows. {Note first that $C(t)\leq K$ for every $t$. Moreover, if $a$ is such that $\phi(a)>K$, then $B(t)>K$ and $C(t)=0$.} 
\rev{Using the upper bounds \eqref{recursive-time-gen} on the time spent by the process $A(t)$ in the various states,} we conclude
$$\ba{rcl}\E_{\0}[J] &\leq &K\Bigg(\E[T_0|X(0)=\0]\\&&\qquad+\ds\sum\limits_{a=1}^{n-1}\E[T_a|X(0)=\0]{{\mathbbm 1}}_{[0,K]}\big(\phi(a)\big)\Bigg)\\[10pt]
&=&K\left( \ds\frac{\beta}{2\beta -1}\frac{1}{\phi_K(0)} +\frac{1}{2\beta -1}\ds\sum\limits_{a=1}^{n-1}\frac{{{\mathbbm 1}}_{[0,K]}\big(\phi(a)\big)}{\phi_K(a)}\right)\\[10pt]
&=&\ds\frac{\beta}{2\beta-1}+\frac{\left|\left\{1\leq a<n\,:\,\phi(a)<K\right\}\right|}{2\beta-1}.\ea$$
%
%
\end{IEEEproof}\smallskip

We observe that the two bounds in Corollary~\ref{cor:feedback} depend on the control policy $\nu$ through the {parameter} $K$ and on the network structure through the minimum conductance profile $\phi$. Due to the monotonicity of $\phi_K$, the upper bound in~\eqref{time control} is  nonincreasing in $K$, while~\eqref{cost control} is nondecreasing in $K$. This {establishes} a trade-off {behavior} between faster spread and higher cost (see, e.g., Figure \ref{fig:sbm_cost}), paving the way for the {formalization} of an optimization problem to choose the value $K$ that best compromises fast spread (larger values of $K$) and affordable cost (lower values of $K$).


\subsection{Fast spread on SBMs}

We consider the feedback control policy in \eqref{feedback-control} on the SBMs introduced in Section~\ref{SBM}, choosing $K<c\alpha p/2$. From~\eqref{eq:expansiveness},  we have that 
$$\phi(a)\geq K\quad\forall\,a\in\{1,\dots , n-1\}\setminus\{n_1,n_2\}\,,$$
with high probability as $n$ grows large. Therefore,
$$\ba{lllll}\phi_K(a)&=&\phi(a)&\text{for}& a\in\{1,\dots , n-1\}\setminus\{n_1,n_2\},\\ \phi_K(a)&\geq& K&\text{for}& a\in\{n_1,n_2\}.\ea$$
{The application of \eqref{time control} yields the following bound on the expected spreading time}
\be\label{eq:boundSBM}\E_{\0}[T]\leq \frac{\beta}{(2\beta-1)K}+\frac{1}{2\beta-1}\left(\frac{2}{K}+\frac{12}{c\alpha p}(\ln(n/2)+1) \right),
\ee
{where the algebraic computations follow the ones carried on in Section~\ref{SBM}}. Using~\eqref{cost control}, we finally bound the expected control cost as
$$\E_{\0}[J]\leq \ds\frac{2+\beta}{2\beta-1}.$$
Hence, with a bounded control cost, we achieve an expected spreading time that grows logarithmically in $n$.

Figure~\ref{fig:sbm} compares our feedback control policy with constant control policies, highlighting the improvement in performance of the former with respect to the latter. We also notice that, for highly connected communities (i.e., for large values of $p$), such an improvement increases in magnitude. 
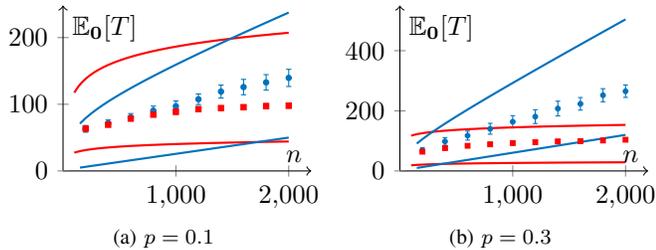
\begin{figure}
\centering
\subfloat[$p=0.1$]{\definecolor{mycolor1}{rgb}{0.00000,0.44700,0.74100}%
\begin{tikzpicture}
\begin{axis}[%
 axis lines=middle,
    axis line style={->},width=\l  cm,
height=\h cm,
at={(0.741in,0.457in)},
scale only axis,
xmin=0,
xmax=2200,
xlabel={$n$},
xtick={0,1000,2000},
ymin=0,
ymax=248,
ylabel={$\E_{\0}[T]$},
axis background/.style={fill=white},
 extra y ticks ={0},
    extra y tick labels={$0$},
]

\addplot [only marks, mark size=1, color=mycolor1,solid,forget plot] 
   plot [error bars/.cd, y dir = both, y explicit]
 table[row sep=crcr, y error plus index=2, y error minus index=2]{%
200	63.1469255513453	5.07174993626754	4507174993626754\\
400	71.4013968487594	5.10459062752116	5.10459062752116\\
600	80.4221188924956	5.64001666857414	5.64001666857414\\
800	90.4370773562378	6.7826441375093	6.7826441375093\\
1000	97.0888383673488	7.91475650056828	7.91475650056828\\
1200	107.524592282668	8.00656212490333	8.00656212490333\\
1400	119.039277070325	9.7197478034905	9.7197478034905\\
1600	125.826005190907	11.5274517664391	11.5274517664391\\
1800	132.838194773249	11.6098955192166	11.6098955192166\\
2000	139.557838351512	12.8896960347816	12.8896960347816\\
};

\addplot [only marks, mark size=1, color=red, mark=square*,solid,forget plot] 
   plot [error bars/.cd, y dir = both, y explicit]
 table[row sep=crcr, y error plus index=2, y error minus index=2]{%
200	63.4003639603731	1.99397413220878	1.99397413220878\\
400	69.2663541227761	2.05989264374623	2.05989264374623\\
600	78.5855917212932	2.25276491609374	2.25276491609374\\
800	84.3835494441192	3.06703776349614	3.06703776349614\\
1000	88.7677577570846	3.84306946206786	3.84306946206786\\
1200	92.7594501863374	3.14890535517177	3.14890535517177\\
1400	94.3407746903203	3.17724049525038	3.17724049525038\\
1600	95.6102590151396	3.53162969957728	3.53162969957728\\
1800	97.180478736256	3.9198746253927	3.9198746253927\\
2000	98.0026375267406	3.96196320850716	3.96196320850716\\
};

\addplot [color=red,solid, thick]
  table[row sep=crcr]{%
100	117.360690162844\\
120	122.830336866663\\
140	127.454857261481\\
160	131.460799040216\\
180	134.994290109908\\
200	138.155105579643\\
220	141.014410973773\\
240	143.624752283461\\
260	146.026033513667\\
280	148.249272678279\\
300	150.319058822888\\
320	152.255214457015\\
340	154.073953111508\\
360	155.788705526706\\
380	157.410722164815\\
400	158.949520996441\\
420	160.413225921524\\
440	161.808826390571\\
460	163.142379267696\\
480	164.41916770026\\
500	165.643827535867\\
520	166.820448930466\\
540	167.952658769951\\
560	169.043688095077\\
580	170.096427689416\\
600	171.113474239686\\
620	172.097168924376\\
640	173.049629873813\\
660	173.972779633816\\
680	174.868368528306\\
700	175.737994634504\\
720	176.583120943505\\
740	177.405090169148\\
760	178.205137581613\\
780	178.984402173711\\
800	179.743936413239\\
820	180.484714790951\\
840	181.207641338322\\
860	181.913556260628\\
880	182.603241807369\\
900	183.277427482931\\
920	183.936794684494\\
940	184.581980841123\\
960	185.213583117058\\
980	185.83216173314\\
1000	186.438242952666\\
1020	187.032321771551\\
1040	187.614864347264\\
1060	188.186310196385\\
1080	188.74707418675\\
1100	189.297548346795\\
1120	189.838103511876\\
1140	190.369090824858\\
1160	190.890843106214\\
1180	191.403676106993\\
1200	191.907889656484\\
1220	192.403768715021\\
1240	192.891584341174\\
1260	193.371594581567\\
1280	193.844045290611\\
1300	194.30917088669\\
1320	194.767195050614\\
1340	195.21833137155\\
1360	195.662783945105\\
1380	196.100747927739\\
1400	196.532410051302\\
1420	196.957949101061\\
1440	197.377536360303\\
1460	197.791336024273\\
1480	198.199505585946\\
1500	198.602196195911\\
1520	198.999552998411\\
1540	199.391715445432\\
1560	199.778817590509\\
1580	200.160988363832\\
1600	200.538351830038\\
1620	200.911027429995\\
1640	201.279130207749\\
1660	201.642771023719\\
1680	202.002056755121\\
1700	202.357090484531\\
1720	202.707971677427\\
1740	203.054796349459\\
1760	203.397657224168\\
1780	203.736643881786\\
1800	204.071842899729\\
1820	204.403337985327\\
1840	204.731210101293\\
1860	205.055537584419\\
1880	205.376396257921\\
1900	205.693859537838\\
1920	206.007998533856\\
1940	206.318882144923\\
1960	206.626577149939\\
1980	206.931148293859\\
2000	207.232658369464\\
};

\addplot [color=red,solid, thick]
  table[row sep=crcr]{%
100	27.387154308503\\
120	28.4234856359385\\
140	29.2931331299537\\
160	30.0494928185725\\
180	30.7154183806849\\
200	31.308115863379\\
220	31.8458298289533\\
240	32.3361697581361\\
260	32.785557614844\\
280	33.2025849546875\\
300	33.5905313412815\\
320	33.9523623869281\\
340	34.2929052973921\\
360	34.6138000166764\\
380	34.9166039696133\\
400	35.2043493853256\\
420	35.4779377103156\\
440	35.7382591549927\\
460	35.987370128625\\
480	36.225798501981\\
500	36.454083779788\\
520	36.673700541998\\
540	36.8849713643562\\
560	37.088236950939\\
580	37.2846003389096\\
600	37.4742641155332\\
620	37.6574496674807\\
640	37.8350098795546\\
660	38.0070738241375\\
680	38.1737880233889\\
700	38.3358297074404\\
720	38.4932812238466\\
740	38.6462404404953\\
760	38.7952569548713\\
780	38.9403825501055\\
800	39.0816825704885\\
820	39.2196112096917\\
840	39.3541998031421\\
860	39.4854914561267\\
880	39.6138673286051\\
900	39.7393448390582\\
920	39.8619516076918\\
940	39.9820117372255\\
960	40.0995330546238\\
980	40.2145322424071\\
1000	40.3272880390369\\
1020	40.4378016071768\\
1040	40.5460818191044\\
1060	40.6523707906123\\
1080	40.7566650372949\\
1100	40.8589678114542\\
1120	40.9594912639578\\
1140	41.0582286747787\\
1160	41.1551792342016\\
1180	41.2505302826741\\
1200	41.3442728647055\\
1220	41.4364032315975\\
1240	41.5270879626589\\
1260	41.6163165830135\\
1280	41.7040832237296\\
1300	41.7905369257719\\
1320	41.8756662117646\\
1340	41.9594636951871\\
1360	42.0420634739283\\
1380	42.1234534434769\\
1400	42.2036251468283\\
1420	42.2826998510182\\
1440	42.3606650980478\\
1460	42.4375116942556\\
1480	42.5133498115566\\
1500	42.5881668382705\\
1520	42.6619530945344\\
1540	42.7348090960921\\
1560	42.8067222233131\\
1580	42.877682498654\\
1600	42.9477819842472\\
1620	43.0170081583508\\
1640	43.085350887872\\
1660	43.1528947938054\\
1680	43.2196275285258\\
1700	43.2855389105228\\
1720	43.3507069779293\\
1740	43.4151196113738\\
1760	43.4787666614908\\
1780	43.5417203157234\\
1800	43.6039687205535\\
1820	43.665501819038\\
1840	43.7263865762006\\
1860	43.7866114295607\\
1880	43.8461664593265\\
1900	43.9051139501587\\
1920	43.9634426464716\\
1940	44.0211427983842\\
1960	44.0782724802545\\
1980	44.1348207522438\\
2000	44.1907780579078\\
};

\addplot [color=mycolor1,solid,forget plot,thick]
  table[row sep=crcr]{%
150	71.0586489742685\\
200	79.1866835151315\\
250	86.239076037035\\
300	92.6111019836009\\
350	98.5136133140552\\
400	104.072469857797\\
450	109.368853785404\\
500	114.458195713034\\
550	119.380032043106\\
600	124.163554992933\\
650	128.830933454159\\
700	133.399399656721\\
750	137.882614181504\\
800	142.291589533796\\
850	146.635333230737\\
900	150.921306794736\\
950	155.155760482574\\
1000	159.3439820557\\
1050	163.490484791857\\
1100	167.599151719105\\
1150	171.673347761953\\
1200	175.716008002266\\
1250	179.729707910937\\
1300	183.716719796825\\
1350	187.679058596539\\
1400	191.61851933272\\
1450	195.536707996241\\
1500	199.435067190836\\
1550	203.314897571219\\
1600	207.177375876462\\
1650	211.023570187575\\
1700	214.854452906736\\
1750	218.670911854624\\
1800	222.473759804068\\
1850	226.263742707204\\
1900	230.04154682524\\
1950	233.807804931961\\
2000	237.563101731699\\
};
\addplot [color=mycolor1,solid,forget plot,thick]
  table[row sep=crcr]{%
150	5\\
2000	50\\
};
\end{axis}
\end{tikzpicture}%
}\,\,\subfloat[$p=0.3$]{\definecolor{mycolor1}{rgb}{0.00000,0.44700,0.74100}%
\begin{tikzpicture}
\begin{axis}[%
 axis lines=middle,
    axis line style={->},width=\l  cm,
height=\h cm,
at={(0.741in,0.457in)},
scale only axis,
xmin=0,
xmax=2200,
xlabel={$n$},
xtick={0,1000,2000},
ymin=0,
ymax=550,
ylabel={$\E_{\0}[T]$},
axis background/.style={fill=white},
 extra y ticks ={0},
    extra y tick labels={$0$},
]

\addplot [only marks, mark size=0.9, color=mycolor1,solid,forget plot] 
   plot [error bars/.cd, y dir = both, y explicit]
 table[row sep=crcr, y error plus index=2, y error minus index=2]{%
200	68.0234918432033	9.59516705253561	9.59516705253561\\
400	98.210098521369	12.7258485179532	12.7258485179532\\
600	117.476881147254	14.9681326028057	14.9681326028057\\
800	140.237487314037	18.1643351967695	18.1643351967695\\
1000	163.604516929381	20.2921138961486	20.2921138961486\\
1200	180.676820661436	22.9707927448385	22.9707927448385\\
1400	207.556514575639	24.5307971416595	24.5307971416595\\
1600	223.288046553103	20.9530134883363	20.9530134883363\\
1800	251.895640910938	21.8218585271794	21.8218585271794\\
2000	265.251489259813	20.6738371389374	20.6738371389374\\
};

\addplot [only marks, mark size=0.9, color=red, mark=square*,solid,forget plot] 
   plot [error bars/.cd, y dir = both, y explicit]
 table[row sep=crcr, y error plus index=2, y error minus index=2]{%
200	64.0453453243807	4.67483283845136	4.67483283845136\\
400	76.048905445019	4.98061782841848	4.98061782841848\\
600	83.6642588308789	4.95055650867051	4.95055650867051\\
800	88.6130762130871	4.54187067661717	4.54187067661717\\
1000	92.5845818336498	4.36750493656649	4.36750493656649\\
1200	95.7815943794385	4.0335045472576	4.0335045472576\\
1400	98.2774084027619	4.18543750790488	4.18543750790488\\
1600	99.0792713394445	5.09037929097616	5.09037929097616\\
1800	101.256679582106	5.3018306163641	5.3018306163641\\
2000	103.786507959004	5.44504434969904	5.44504434969904\\
};
\addplot [color=red,solid, thick]
  table[row sep=crcr]{%
100	18.7594193175569\\
150	20.1469260973828\\
200	21.126247825563\\
250	21.8833322633741\\
300	22.5004649876219\\
350	23.021331461431\\
400	23.471913543817\\
450	23.8689219403472\\
500	24.2237398949962\\
550	24.5444707790638\\
600	24.8370879302766\\
650	25.1061211171387\\
700	25.3550870424517\\
750	25.5867704833251\\
800	25.8034142261484\\
850	26.0068511420145\\
900	26.1985983402113\\
950	26.3799257568049\\
1000	26.5519070793384\\
1050	26.7154581992307\\
1100	26.8713666863603\\
1150	27.0203146889787\\
1200	27.1628969436609\\
1250	27.2996350969824\\
1300	27.430989209632\\
1350	27.5573670829433\\
1400	27.6791318844122\\
1450	27.796608431379\\
1500	27.9100884065731\\
1550	28.0198347162319\\
1600	28.1260851545558\\
1650	28.2290555028946\\
1700	28.3289421651679\\
1750	28.425924420373\\
1800	28.520166357055\\
1850	28.6118185421446\\
1900	28.701019466758\\
1950	28.7878968038018\\
2000	28.8725685060279\\
};

\addplot [color=red, solid,thick]
  table[row sep=crcr]{%
100	117.277609398471\\
120	119.465468079999\\
140	121.315276237926\\
160	122.91765294942\\
180	124.331049377297\\
200	125.59537556519\\
220	126.739097722842\\
240	127.783234246718\\
260	128.7437467388\\
280	129.633042404645\\
300	130.460956862488\\
320	131.235419116139\\
340	131.962914577936\\
360	132.648815544016\\
380	133.297622199259\\
400	133.91314173191\\
420	134.498623701943\\
440	135.056863889562\\
460	135.590285040412\\
480	136.101000413437\\
500	136.59086434768\\
520	137.06151290552\\
540	137.514396841314\\
560	137.950808571364\\
580	138.3719044091\\
600	138.778723029208\\
620	139.172200903084\\
640	139.553185282859\\
660	139.92244518686\\
680	140.280680744656\\
700	140.628531187135\\
720	140.966581710735\\
740	141.295369400993\\
760	141.615388365979\\
780	141.927094202818\\
800	142.230907898629\\
820	142.527219249714\\
840	142.816389868662\\
860	143.098755837585\\
880	143.374630056281\\
900	143.644304326506\\
920	143.908051207131\\
940	144.166125669783\\
960	144.418766580157\\
980	144.666198026589\\
1000	144.9086305144\\
1020	145.146262041954\\
1040	145.379279072239\\
1060	145.607857411887\\
1080	145.832163008033\\
1100	146.052352672052\\
1120	146.268574738084\\
1140	146.480969663276\\
1160	146.689670575819\\
1180	146.894803776131\\
1200	147.096489195927\\
1220	147.294840819342\\
1240	147.489967069803\\
1260	147.68197116596\\
1280	147.870951449578\\
1300	148.05700168801\\
1320	148.240211353579\\
1340	148.420665881953\\
1360	148.598446911375\\
1380	148.773632504429\\
1400	148.946297353854\\
1420	149.116512973758\\
1440	149.284347877455\\
1460	149.449867743043\\
1480	149.613135567712\\
1500	149.774211811698\\
1520	149.933154532698\\
1540	150.090019511506\\
1560	150.244860369537\\
1580	150.397728678866\\
1600	150.548674065348\\
1620	150.697744305331\\
1640	150.844985416433\\
1660	150.990441742821\\
1680	151.134156035382\\
1700	151.276169527146\\
1720	151.416522004304\\
1740	151.555251873117\\
1760	151.692396223\\
1780	151.827990886048\\
1800	151.962070493225\\
1820	152.094668527464\\
1840	152.22581737385\\
1860	152.355548367101\\
1880	152.483891836502\\
1900	152.610877148468\\
1920	152.736532746876\\
1940	152.860886191302\\
1960	152.983964193309\\
1980	153.105792650877\\
2000	153.226396681119\\
};

\addplot [color=mycolor1,solid,forget plot,thick]
  table[row sep=crcr]{%
150	91.0586489742685\\
200	105.853350181798\\
250	119.572409370368\\
300	132.611101983601\\
350	145.180279980722\\
400	157.405803191131\\
450	169.368853785404\\
500	181.124862379701\\
550	192.71336537644\\
600	204.163554992933\\
650	215.497600120826\\
700	226.732732990054\\
750	237.882614181503\\
800	248.958256200463\\
850	259.96866656407\\
900	270.921306794736\\
950	281.822427149241\\
1000	292.677315389033\\
1050	303.490484791857\\
1100	314.265818385772\\
1150	325.006681095286\\
1200	335.716008002266\\
1250	346.396374577603\\
1300	357.050053130158\\
1350	367.679058596539\\
1400	378.285185999387\\
1450	388.870041329575\\
1500	399.435067190836\\
1550	409.981564237886\\
1600	420.510709209795\\
1650	431.023570187575\\
1700	441.521119573403\\
1750	452.004245187957\\
1800	462.473759804068\\
1850	472.93040937387\\
1900	483.374880158573\\
1950	493.807804931961\\
2000	504.229768398366\\
};
\addplot [color=mycolor1,solid,forget plot,thick]
  table[row sep=crcr]{%
150	10\\
2000	120\\
};

\end{axis}
\end{tikzpicture}%
}
\caption{Monte Carlo estimation ($200$ simulations) and $90\%$ confidence intervals of the expected spreading time $\E_{\0}[T]$ on SBMs for different $n$ with  $\alpha=1$, $\beta=0.8$, $c=0.4$, and $L=5$, under the feedback control policy in~\eqref{feedback-control} with $K=1/4$ (red squares) and constant control (blue circles), for two different values of $p$. \rev{The red solid curves are a numerical estimation of~\eqref{eq:minimization time2} and the theoretical bound for the feedback control from~\eqref{eq:boundSBM}. The blue solid curves are theoretical bounds for the constant control from~\eqref{SBM-linear} and~\eqref{SBM-linear2}.}
}
\label{fig:sbm}
\end{figure}

In Figure \ref{fig:sbm_cost}, we analyze the trade-off between spreading time and control cost depending on the parameter $K$. There is a first phase, for $K\leq c\alpha p/2$ (light blue), where, {as $K$ increases, the average spreading time shows a fast decrease, while the average control cost increases slowly. Then, for $K>c\alpha p/2$ (dark blue), the average spreading time decreases slowly against a strong growth of the control cost}. In the same figure, we compare our control policy with a simpler feedback control policy that does not depend on $B(t)$, obtained using the same target function $\iota(X(t))$ but a non-negative constant rate {(different squares in green correspond to different values)}. This comparison supports our intuition that the use of $B(t)$ in the feedback law {is key to improve in the performance of the spreading process.}
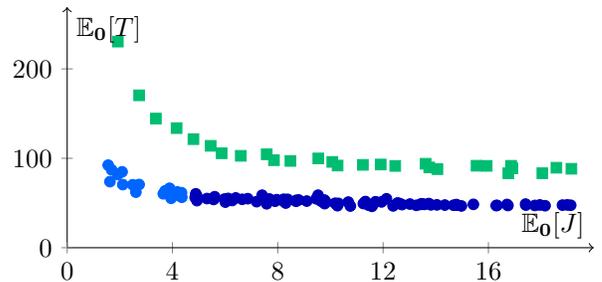
\begin{figure}
\centering
\definecolor{mycolor1}{RGB}{0,100,255}%
\definecolor{mycolor3}{RGB}{0,0,185}%
\definecolor{mycolor2}{rgb}{0.00000,0.74100,0.44700}%
\begin{tikzpicture}
\begin{axis}[%
 axis lines=middle,
    axis line style={->},width=7 cm,
height=3.2 cm,
at={(0.741in,0.457in)},
scale only axis,
xmin=0,
xmax=20,
xlabel={$\E_{\0}[J]$},
xtick={0,4,8,12,16},
ymin=0,
ymax=269,
ylabel={$\E_{\0}[T]$},
axis background/.style={fill=white},
 extra y ticks ={0},
    extra y tick labels={$0$},
    ylabel shift = -20 pt,
     extra x ticks ={0},
    extra x tick labels={$0$},
]


\addplot [only marks,color=mycolor1,solid,forget plot]
  table[row sep=crcr]{%
1.55861170101457	92.3319454959756\\
1.68766957819245	87.1572171412387\\
2.08651642087591	84.8945559655958\\
1.92086121558134	81.8437019166012\\
1.62470755312959	73.9023330448968\\
2.11267890234148	70.5739169685292\\
2.49002059627056	70.3950150229896\\
2.73403192394754	70.6794816872263\\
2.6113653939888	62.1296389547977\\
3.7202230270828	63.5789943109776\\
3.89015571184104	66.4827884005505\\
3.65153455209073	60.325418473595\\
4.3562142426707	56.4066562126847\\
3.94794564106521	55.2681369966731\\
4.26355069922907	58.0198434041267\\
3.75121640689276	62.8136658402524\\
4.18657778963389	62.4073958947977\\
4.34490754119249	61.5166830777683\\
};

\addplot [only marks,color=mycolor3,solid,forget plot]
  table[row sep=crcr]{%
5.60214852642527	57.040575869457\\
4.93094229774562	52.6860358749879\\
5.97469846541258	54.3949601534791\\
6.1130896895464	55.1182287583307\\
6.00919732015812	51.0112084893822\\
7.40562120147685	58.7216674003876\\
8.22429499768756	54.2198569317582\\
8.22171684283995	52.0033111026953\\
7.19684535208603	51.499258743315\\
8.37234870798495	54.1226704093386\\
7.57659887931731	49.1249714951194\\
9.53063668235484	58.7523723712008\\
9.46059287146793	54.7151196235934\\
8.69454632682131	51.9217538330084\\
6.64788128457716	54.150363941453\\
9.78204787226354	49.3157554152001\\
9.77616723111502	53.2652557697061\\
10.3049531874229	49.5854715713646\\
11.3298716058629	51.6016931571492\\
12.1285660676272	54.6916390586486\\
11.7133864868948	51.412444639737\\
10.2665550362279	46.9920007714809\\
11.5571036069361	47.0837139296605\\
11.4450758329629	47.6400107569136\\
12.6211948360476	49.474213644212\\
10.7570127331913	46.6328459766202\\
12.3785729313519	46.7780850013279\\
11.5762038365177	46.2956790852097\\
13.7353230442555	49.4111889600146\\
13.5699297288584	48.4515479952616\\
13.2614015024848	47.393025919211\\
14.3454566415613	47.6921097488428\\
13.815258530609	47.9369256118135\\
16.2944985479555	47.1489510620901\\
17.4249247318581	47.6862220349121\\
4.88568191088998	60.2835248304906\\
4.92821569908782	57.6995642908145\\
4.8615494969012	56.1816194800228\\
5.31143882163055	54.9877574389875\\
5.58691190102206	54.8493373517981\\
5.56091583323224	53.9787266246026\\
6.40278900615778	55.6652497435744\\
6.26839579346628	52.808029922329\\
6.8605882361411	54.8556138205917\\
7.68383986993039	54.4853331892568\\
7.96115533118644	53.7283406728199\\
7.89853241511284	51.987802559337\\
8.70187977962721	52.9829313944454\\
8.79371027118908	53.408661351266\\
8.30266925337313	49.9384396963676\\
9.63975717614387	52.1787279294428\\
9.10197812212466	52.152888912996\\
9.82427224441883	51.0485224328611\\
9.72942758965521	51.0042689513245\\
10.1783450293646	49.7327768805485\\
10.6780436442312	51.0299040822672\\
11.3252280214719	51.0778819417514\\
11.9415043936627	51.2332360256357\\
11.2507102964984	49.3906462633237\\
12.5478776483152	50.0098837994135\\
11.6186671721724	48.3705265816324\\
12.7201486846479	48.1943049144957\\
13.4630210596933	49.1357005488085\\
13.0187177067138	48.8146814183674\\
13.4021818604659	48.3348735495994\\
13.4820266566211	48.843593892779\\
14.0438515520921	47.9225632106569\\
14.7861216320319	48.2624967165433\\
15.430224205024	48.492972392409\\
14.6775888252317	47.514157969173\\
14.787855638664	47.6559435542091\\
14.9746510642691	46.7256170186737\\
16.7068653183332	48.3629430515721\\
16.3271669584273	47.3213425314659\\
16.7426883966302	47.2172455793398\\
17.9535640242645	47.8565569433558\\
17.7646422035568	47.2677089296831\\
17.4131797439771	48.7819528920852\\
18.1525936417161	46.9681665716806\\
19.1257445312928	47.4815499867461\\
18.9931536427963	47.8782438942652\\
18.8131893326942	47.6362462431305\\
};

\addplot [only marks,mark=square*,color=mycolor2,solid,forget plot]
  table[row sep=crcr]{%
1.92558320585359	230.505961174313\\
2.73466064364166	170.437004467967\\
3.3698190680151	144.425336953857\\
4.15690638722464	133.772588788002\\
4.7968223263288	121.431037459938\\
5.45093534637108	113.942724518746\\
5.86758586327485	105.638678870449\\
6.5949491924471	102.771849462506\\
7.57927146299442	104.446666867686\\
7.85567167372078	97.8904797749627\\
8.48038881206308	97.0426057307196\\
9.54049998131696	99.8279426026488\\
10.0710928115561	95.9701876255707\\
10.2724385667897	91.8830690455917\\
11.2330709784619	92.6591910816466\\
11.9094387268818	93.0569070986051\\
12.4613271708584	91.457881061735\\
13.6158585366198	93.9205454106511\\
13.7525702923972	89.7779405551123\\
14.0627872633082	87.9043164043584\\
15.5514622727559	91.6938466602628\\
15.9356076345489	91.5300348430771\\
16.8618792028691	91.6283016836889\\
16.9129521759067	88.7641557350773\\
16.7574676396485	83.2799809737871\\
18.5827834425121	89.4595548781183\\
18.0491501844354	83.3177051994508\\
19.1521259275198	88.2948997402741\\
};

\end{axis}
\end{tikzpicture}%
\caption{Monte Carlo estimation ($200$ simulations) of the expected spreading time $\E_{\0}[T]$ and expected cost $\E_{\0}[J]$ on SBMs with $n=800$, $\alpha=1$, $\beta=0.8$, $c=0.4$, $p=0.1$ and $L=5$, using the feedback control policy in \eqref{feedback-control}, for different values of the parameter $K$. Green squares are Monte Carlo estimation of the expected spreading time $\E_{\0}[T]$ and expected cost $\E_{\0}[J]$ under a control policy with the same position , but constant rate.}
\label{fig:sbm_cost}
\end{figure}

Summing up, despite (i) the myopic choice of our control policy in which no optimization is done on the target node, (ii) the limited a-priori knowledge on the topology (only knowledge of $\phi$ is used), and (iii) the use of only two one-dimensional statistics (from an $n$-dimensional state variable), we have been able to design a feedback control policy whose performance is guaranteed by analytical results and whose effectiveness (already numerically tested on a case study in~\cite{iscas2018}) has been explicitly demonstrated on SBMs. We remark that  the control strategy is designed without knowledge of the exact partition of nodes into communities, and even of their size. This property provides robustness in real-world situations, where limited data might be available leading to uncertainty on the exact network structure.

\section{Discussion and Conclusion}\label{section conclusions}

We have proposed a \rev{continuous-time stochastic} dynamical model that allows for an analytical treatment of evolutionary dynamics and incorporates exogenous control input mechanisms. 
The main contributions of this paper are i) the formulation of a link-based spreading process incorporating an exogenous control and leading to a non-homogenous Markov process; ii) a rigorous analysis of the transient spreading process in terms of the network topology and of the control policy adopted; iii) the derivation of a set of ready-to-use corollaries \rev{to estimate the spreading time and the control cost for specific control policies}; and iv) the design and the analysis of an effective feedback control policy, which relies on few topological data and one-dimensional variables of the system. 
The proposed policy provides a dramatic improvement in the speed of the evolutionary process for some network structures where simpler control policies fail to achieve fast spread. 

{ As a consequence of our analysis, we have characterized three classes of network topologies depending on their controllability.\begin{enumerate}
\item[a)]
{\bf Topologies easy to control even with constant control policies.} Among them we mention expander graphs, which is a large family of  highly connected networks {including} the complete graph, Erd\H{o}s-R\'enyi random graphs, small-world networks{, as well many scale-free networks such as those generated according to a preferential attachment algorithm}. For these topologies, little effort is required to achieve fast spread, that is, to guarantee that the expected spreading time grows logarithmically with the network size $n$ for every constant control policy.
\item[b)]
{\bf Topologies easy to control only with feedback control policies.} A representative example is the stochastic block model (SBM), of which barbell graphs are well known limit cases. For these structures, no fundamental limit precludes fast spread. However, we have shown that constant control policies fail in achieving it. A major contribution of this paper then consisted in the development of a feedback control policy that remarkably improves the speed of the  process, achieving fast spread.
\item[c)]
{\bf Topologies hard to control under any control policy}, of which the ring graph is a representative element. For these topologies, we observe slow spread under any feasible control policy, where the expected spreading time shows a linear growth in the network size.
\end{enumerate}}



We believe that the generality of Theorem~\ref{teo:main upper} and the effectiveness of the feedback control policy proposed in Section~\ref{sec:feedback} pave the way for further research seeking for an optimal control policy for evolutionary dynamics, also extending the real-world case study in \cite{iscas2018}, \rev{in which evolutionary dynamics are proposed to model the introduction of genetically modified mosquitoes in a geographic region to substitute the species that transmit diseases~\cite{Buckman2020}. The proposed feedback control policy requires limited a-priori knowledge of the network topology. When more information on the network structure is available, we believe that such information might be exploited to design and study targeted introduction policies. For instance,  inspired by targeted vaccination strategies~\cite{Chung2009}, the introduction of the novel states might be prioritized in nodes with high network centrality.} Other avenues of future research include the use of the technical tools developed in this paper to study control policies on other dynamical processes on networks, such as opinion dynamics, diffusion of information, and epidemics.

\appendix
\subsection{Proof of Lemma \ref{prop:Moran}}\label{section analysis}

For a given time $t_0\in\R_+$  and let $X_k$, $A_k$, $B_k$, $C_k$ be the discrete-time jump processes of $X(t)$, $A(t)$, $B(t)$, and $C(t)$, respectively, starting from time $t^+_0$ and onwards, with $X_0=X(t_0)=\x$. Clearly,
\be\label{confront1}\P[A(t)\geq a,\,t\geq t_0\,|\, X(t_0)=\x]=\P[A_k\geq a,\,k\ge0\,|\, X_0=\x]\ee
It follows from \eqref{eq:l1}--\eqref{eq:l2} that  the increase and decrease transition probabilities of $A_k$ conditioned to the process $X_k$ at time $k$, also satisfy
\be
p_k^-(a|\x)\leq 1-\beta,\quad p_k^+(a|\x)\geq \beta,\quad\text{for }a\notin\{0,n\},
\ee
while $p_k^+(0|0)=1$ for all $k$, and $a=n$ is an absorbing state.

Consider now a discrete-time birth-and-death chain $\tilde A_k$ with state space $\{0,\dots ,n\}$ and transition probabilities 
$$\left\{\begin{array}{l}p^+(0)=1,\\p^+(a)=\beta,\,a\notin\{0,n\},\\ p^+(n)=0;\end{array}\right.\left\{\begin{array}{l}p^-(0)=0,\\p^-(a)=1-\beta,\,a\notin\{0,n\},\\ p^-(n)=0,\end{array}\right.$$
with $\tilde A_0=a$.
A standard argument allows us to couple the two processes $A_k$ and $\tilde A_k$ in such a way that $A_k\geq \tilde A_k$ for every $k$, yielding 
{\be\label{confront2}\P[A_k\geq a,\, k\ge0\,|\, X_0=\x]\geq \P[\tilde A_k\geq a,\,k\ge0]\,.\ee }
On the other hand, a direct computation for the birth-and-death chain $\tilde A_k$ implies that
{\be\label{confront3}\P[\tilde A_k\geq a,\, \forall \,k]=\ds \frac{1-\left(\frac{1-\beta}{\beta}\right)}{1-\left(\frac{1-\beta}{\beta}\right)^{n-a+1}}\geq\frac{2\beta-1}{\beta}\,.\ee
Considering that $t_0$ can be chosen arbitrarily, \eqref{confront1}, \eqref{confront2}, and \eqref{confront3} prove Lemma  \ref{prop:Moran}. }}\qed

\subsection{Proof of  Lemma \ref{lemma:monotonicity}}\label{app:lemma1}
First, we consider the case $\beta=\gamma$ and $\x_0\leq \y_0$. We define the coupled process $Z(t)=(X(t),Y(t))$  on the state space $\{0,1\}^n\times\{0,1\}^n$, with initial condition $Z(0)=(\x_0,\y_0)$, associated with the same graph $\mc G$. The coupling mechanism is the following. Each link $\{i,j\}$ is equipped with an independent rate-$W_{ij}$ Poisson clock. When the clock associated with link $\{i,j\}$ ticks, the spreading mechanism acts on that link for both $X(t)$ and $Y(t)$ as for a standard controlled evolutionary dynamics and, if a conflict occurs in both processes, then the outcome is the same. Each node $i$  is equipped with a non-homogeneous Poisson clock with rate $U_i(t)$ associated with the external control in node $i$. When the clock associated with node $i$ ticks, then both $X_i$ and $Y_i$ turn to $1$. Each of the two marginals of $X(t)$ and $Y(t)$ coincides with the distribution of a controlled evolutionary dynamics $(\mc G,\beta,U(t))$ with initial condition $X(0)=\x_0$ and $Y(0)=\y_0$, respectively. 

We now show that, under this coupling, $Y(t)\geq X(t)$, for every $t\geq 0$. At $t=0$, this is verified by assumption. In the following, we prove that any transition of $Z(t)$ keeps the inequality preserved. We assume that  a transition of the process occurs at time $t$. If $X(t^-)=Y(t^-)$, then, due to the coupling mechanism, $X(t^+)=Y(t^+)$. If $Y(t^-)>X(t^-)$, we analyze all the possible events which trigger a transition.
\begin{itemize}
\item \emph{Spreading mechanism}. When link $\{i,j\}$ activates and a conflict occurs: 
\begin{enumerate}
\item[(a)] if $X_i(t^-)=Y_i(t^-)=X_j(t^-)=0$, $Y_j(t^-)=1$ and state $1$ wins, then $X_i(t^+)=X_j(t^+)=0$ and $Y_i(t^+)=Y_j(t^+)=1$, so that $Y(t^+)>X(t^+)$; 
\item[(b)] if $X_i(t^-)=Y_i(t^-)=X_j(t^-)=0$, $Y_j(t^-)=1$ and state $0$ wins, then $Y_i(t^+)=Y_j(t^+)=X_i(t^+)=X_j(t^+)=0$, so that $Y(t^+)\geq X(t^+)$; 
\item[(c)] if  $X_i(t^-)=Y_i(t^-)=0$, $X_j(t^-)=Y_j(t^-)=1$ and state $1$  wins, then, $Y_i(t^+)=Y_j(t^+)=X_i(t^+)=X_j(t^+)=1$, so that $Y(t^+)>X(t^+)$;
\item[(d)] if $X_i(t^-)=Y_i(t^-)=0$, $X_j(t^-)=Y_j(t^-)=1$, and state $0$ wins, then, $Y_i(t^+)=Y_j(t^+)=X_i(t^+)=X_j(t^+)=0$, so that $Y(t^+)>X(t^+)$;
\item[(e)] if $X_i(t^-)=0$, $Y_i(t^-)=X_j(t^-)=Y_j(t^-)=1$, and state $1$  wins, then, $Y_i(t^+)=Y_j(t^+)=X_i(t^+)=X_j(t^+)=1$, so that $Y(t^+)\geq X(t^+)$;
\item[(f)] if $X_i(t^-)=0$, $Y_i(t^-)=X_j(t^-)=Y_j(t^-)=1$, and state $0$ wins, then, $X_i(t^+)=X_j(t^+)=0$ and  $Y_i(t^+)=Y_j(t^+)=1$, so that $Y(t^+)> X(t^+)$.
\end{enumerate}
\item \emph{External control}. Node $i$ activates and it has state $0$ at least in one of the two processes: 
\begin{enumerate}
\item if $X_i(t^-)=Y_i(t^-)=0$, then, $X_i(t^+)=Y_i(t^+)=1$, so that $Y(t^+)>X(t^+)$;
\item if $X_i(t^-)=0$, $Y_i(t^-)=1$, then, $X_i(t^+)=Y_i(t^+)=1$, so that $Y(t^+)\geq X(t^+)$.
\end{enumerate}
\end{itemize}
Hence, after each transition of the process, the inequality $Y(t)\geq X(t)$ is preserved. 

The proof for $\beta<\gamma$ and $\x_0=\y_0$ follows a similar argument. We define the coupled process $Z(t)=(X(t),Y(t))$ in which each link $\{i,j\}$ is equipped with an independent Poisson clock with rate $W_{ij}$. When the clock associated with link $\{i,j\}$ ticks, the spreading mechanism acts on that link for both $X(t)$ and $Y(t)$. If a conflict occurs in only one of the two processes, then it is solved as in a standard controlled evolutionary dynamics with probability for the novel state to win the conflict equal to $\beta$ for $X(t)$ and $\gamma$ for $Y(t)$, respectively. If the conflict occurs in both processes, then with probability $\beta$ the novel state wins in both $X(t)$ and $Y(t)$, with probability $\gamma-\beta$ it wins only in $Y(t)$, and with probability $1-\gamma$ the novel state loses in both $X(t)$ and $Y(t)$. Each node $i$ is given an non-homogeneous Poisson clock with rate $U_i(t)$. When the clock associated with node $i$ ticks, then both $X_i$ and $Y_i$ turn to $1$. We immediately deduce that the two marginals $X(t)$ and $Y(t)$ are controlled evolutionary dynamics $(\mc G,\beta,U(t))$ and $(\mc G,\gamma,U(t))$, respectively, with the same initial condition $X(0)=Y(0)=\x_0=\y_0$.

Under this coupling, the inequality $Y(t)\geq X(t)$ holds true for every $t\geq 0$. In fact, at $t=0$ it is verified, since $X(0)=Y(0)$. Then, the analysis of all possible transitions and their effect on the inequality is performed similar to above and is omitted due to space constraints. Finally, the case $\beta<\gamma$ and $\x_0<\y_0$ is obtained by combining the two couplings above.

The coupling $Z(t)=(X(t),Y(t))$ such that  $Y(t)\geq X(t)$ for every $t\geq 0$, proves the stochastic domination $Y(t)\succeq X(t)$~\cite{Draief2006}, yielding $\E_{\y_0}[T_Y]\leq\E_{\x_0}[T_X]$. Since $U(t)\geq 0$, we have
$$\ba{rcl}
\E_{\x_0}[J_X]&=&\ds\E\left[\int_{0}^{T_X}U(t)\text{d}t\right]=\ds\int_{0}^{\E_{\x_0}[T_X]}U(t)\text{d}t\\[10pt]
&\geq&\ds\int_{0}^{\E_{\y_0}[T_Y]}U(t)\text{d}t=\E_{\y_0}[J_X]\,,\ea
$$
which completes the proof. \qed

\subsection{Proof of Lemma \ref{lemma:monotonicity2}}\label{app:lemma2}
We define a coupled process $Z(t)=(X(t),Y(t))$ on the state space $\{0,1\}^n\times\{0,1\}^n$, with initial condition $Z(0)=(\x_0,\y_0)$. Here, the marginal distributions of $X(t)$ and $Y(t)$ are controlled evolutionary dynamics $(\mc G,1, U(t))$ and $(\mc G,1,0)$ associated with a the same graph $\mc G$. The coupling mechanism is the following. Each link $\{i,j\}$ of the graph $\mc G$ is equipped with an independent rate-$W_{ij}$ Poisson clock. When the clock associated with link $\{i,j\}$ ticks, the spreading mechanism acts on that link for both $X(t)$ and $Y(t)$, with $\beta=1$.  Each node $i$  is given a non-homogeneous Poisson clock with rate $U_i(t)$, associated with the external control in node $i$. When the clock associated with node $i$ ticks, the state $X_i(t)$ turns to $1$. We immediately deduce that the two marginals of $X(t)$ and $Y(t)$ are controlled evolutionary dynamics $(\mc G,1,U(t))$ and $(\mc G,1,0)$ with the desired initial conditions, respectively.

We show now that, under this coupling, $Y(t)\geq X(t)$,  for every $t$. At $t=0$ this is verified by assumption. In the following, we show that, after each transition of the coupled process, the inequality is preserved. 
\begin{itemize}
\item \emph{Spreading mechanism.} Since $\beta=1$, the novel state always wins. So only three transitions can occur when link $\{i,j\}$ activates and a conflict occurs:
\begin{enumerate}
\item[(a)] If $X_i(t^-)=Y_i(t^-)=X_j(t^-)=0$ and $Y_j(t^-)=1$, then $X_i(t^+)=X_j(t^+)=0$,  $Y_i(t^+)=Y_j(t^+)=1$,  which implies that $Y(t^+)>X(t^+)$.
\item[(b)] If $X_i(t^-)=Y_i(t^-)=0$ and $X_j(t^-)=Y_j(t^-)=1$, then $Y_i(t^+)=Y_j(t^+)=X_i(t^+)=X_j(t^+)=1$, which implies that  $X(t^+)\geq Y(t^+)$.
\item[(c)] If $X_i(t^-)=0$ and $Y_i(t^-)=Y_j(t^-)=X_j(t^-)=1$, then $Y_i(t^+)=Y_j(t^+)=X_i(t^+)=X_j(t^+)=1$,  which implies that $Y(t^+)\geq X(t^+)$.
\end{enumerate}
\item \emph{External control.} We observe that (i) we can only control nodes for which $Y_i(0)=1$, and (ii) since $\beta=1$, if $Y_i(0)=1$ then $Y_i(t)=1$ for every $t\geq 0$. Hence, external control preserves the ordering $Y(t)\geq X(t)$. 
\end{itemize}
The stochastic domination  $Y(t)\succeq X(t)$ yields the claim. \qed

%

\begin{IEEEbiography}
[{\includegraphics[width=1in,height=1.25in,clip,keepaspectratio]{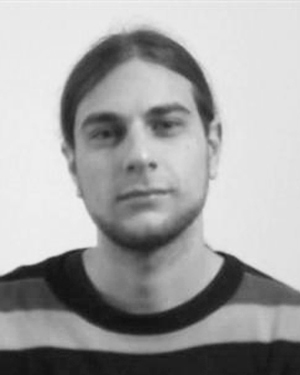}}]
{Lorenzo Zino}
is a PostDoc Researcher at the Faculty of Science and Engineering, University of Groningen, Groningen, The Netherlands, since 2019. He received the Laurea degree (B.Sc.) in Applied Mathematics from Politecnico di Torino, Torino, Italy, in 2012, the Laurea Magistrale degree (M.S.) in Mathematical Engineering from Politecnico di Torino, in 2014, and the Ph.D. in Pure and Applied Mathematics from Politecnico di Torino and Universit\`a di Torino (joint doctorate program), in 2018. He was a Research Fellow at Politecnico di Torino in 2018--19 and a Visiting Research Assistant at New York University Tandon School of Engineering in 2017--18 and in 2019. 
His current research interests include modeling, analysis, and control of dynamical processes over networks, applied probability, network analysis, and game theory.
\end{IEEEbiography}

\begin{IEEEbiography}
[{\includegraphics[width=1in,height=1.25in,clip,keepaspectratio]{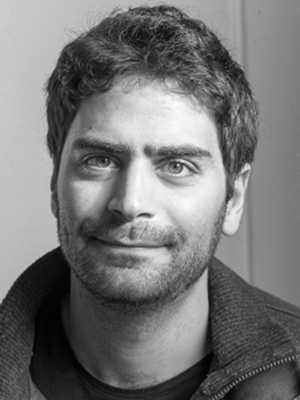}}]
{Giacomo Como}
is  an  Associate  Professor  at  the Department  of  Mathematical  Sciences,  Politecnico di  Torino,  Italy,  and  at  the  Automatic  Control  Department  of  Lund  University,  Sweden.  He  received the B.Sc., M.S., and Ph.D. degrees in Applied Mathematics  from  Politecnico  di  Torino,  in  2002,  2004, and 2008, respectively. He was a Visiting Assistant in  Research  at  Yale  University  in  2006--2007  and  a Postdoctoral  Associate  at  the  Laboratory  for  Information  and  Decision  Systems,  Massachusetts  Institute of Technology, from 2008 to 2011. He currently serves  as  Associate  Editor  of the  \textit{IEEE Transactions on Network Science and Engineering} and of the \textit{IEEE Transactions on Control of Network Systems} and  as  chair  of the  {IEEE-CSS  Technical  Committee  on  Networks  and  Communications}.  He was  the  IPC  chair  of  the  IFAC  Workshop  NecSys'15  and  a  semiplenary speaker  at  the  International  Symposium  MTNS'16.  He  is  recipient  of  the 2015  George S. ~Axelby  Outstanding Paper Award.  His  research interests  are in  dynamics,  information,  and  control  in  network  systems  with  applications to  cyber-physical  systems,  infrastructure  networks,  and  social  and  economic networks
\end{IEEEbiography}

\begin{IEEEbiography}
[{\includegraphics[width=1in,height=1.25in,clip,keepaspectratio]{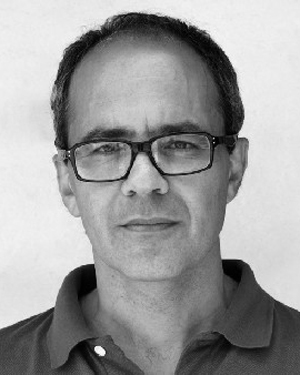}}]
{Fabio Fagnani}
received the Laurea degree in Mathematics from the University of Pisa and the Scuola Normale Superiore, Pisa, Italy, in 1986. He received the PhD degree in Mathematics from the University of Groningen,  Groningen,  The  Netherlands,  in 1991. From 1991 to 1998, he was an Assistant Professor of Mathematical Analysis at the Scuola Normale Superiore. In 1997, he was a Visiting Professor at the Massachusetts Institute of Technology (MIT), Cambridge, MA. Since 1998, he has been with the Politecnico of Torino, where he is currently (since 2002) a Full Professor of Mathematical Analysis. From 2006 to 2012, he has acted as Coordinator of the PhD program in Mathematics for Engineering Sciences at Politecnico di Torino. From June 2012 to September 2019, he served as the Head of the Department of Mathematical Sciences, Politecnico di Torino. His current research topics are on cooperative algorithms and dynamical systems over graphs, inferential distributed algorithms, and opinion dynamics. He is an Associate Editor of the \textit{IEEE Transactions on Automatic Control} and served in the same role for the \textit{IEEE Transactions on Network Science and Engineering} and of the \textit{IEEE Transactions on Control of Network Systems}.
\end{IEEEbiography}




\end{document}